\documentclass[aps,twocolumn,showpacs,showkeys]{revtex4} 

\usepackage{epsf}
\usepackage{dcolumn}
\usepackage{bm}

\begin{document}


\title{Network rigidity at finite temperature: Relationships between
thermodynamic stability, the non-additivity of entropy and cooperativity
in molecular systems}%

\author{Donald J. Jacobs}
\email{Corresponding Author: donald.jacobs@csun.edu}
\homepage{http://www.csun.edu/~dj54698}
\author{S. Dallakyan}
\author{G. G. Wood}
\author{A. Heckathorne}
\affiliation{Physics and Astronomy Department\\
             California State University, Northridge, CA 91330}

\date{\today}

\begin{abstract}
A statistical mechanical distance constraint model (DCM) is presented that explicitly 
accounts for network rigidity among constraints present within a system.
Constraints are characterized by local microscopic free energy functions. 
Topological re-arrangements of thermally fluctuating constraints are
permitted. The partition function is obtained by combining microscopic
free energies of individual constraints using network rigidity as an 
underlying long-range mechanical interaction --- giving a quantitative
explanation for the non-additivity in component entropies exhibited in 
molecular systems. Two exactly solved 2-dimensional toy
models representing flexible molecules that can undergo conformational 
change are presented to elucidate concepts, and to outline a DCM
calculation scheme applicable to many types of physical
systems. It is proposed that network rigidity plays a central role 
in balancing the energetic and entropic contributions to the free
energy of bio-polymers, such as proteins. As a demonstration, 
the distance constraint model is solved exactly for the alpha-helix 
to coil transition in homogeneous peptides. Temperature and size 
independent model parameters are fitted to Monte Carlo simulation data,
which includes peptides of length 10 for gas phase, and lengths 10, 15, 20
and 30 in water. The DCM is compared to the Lifson-Roig model. It is found 
that network rigidity provides a mechanism for cooperativity in molecular
structures including their ability to spontaneously self-organize. In particular, the formation of a characteristic topological arrangement
of constraints is associated with the most probable microstates changing
under different thermodynamic conditions.
\end{abstract}

\maketitle

\section{Introduction\label{sec:level1}}
Network rigidity deals with a system of particles subjected to a set of 
constraints. Depending on the number and position of these constraints,
the system will have a residual number of independent degrees of freedom.
A simple way of characterizing the degree of mechanical stability of a
given framework is to ignore the way constraints are positioned, and
to treat all constraints as independent. In this approximation, the 
number of independent degrees of freedom governing internal motions, $F$, 
in the framework is given by $F = dN - N_c - d(d+1)/2$, where $d$ is the 
dimension of the system, $N$ is the number of vertices, $N_c$ the number
of constraints, and the trivial rigid body motions of the entire framework
subtracted out. The use of constraint counting to determine structural 
stability in macroscopic systems dates back to Maxwell \cite{Maxwell}.
Nearly 25 years ago, 
Philips \cite{Philips} realized that constraint counting is applicable 
to microstructure in covalent glasses by treating central and
bond-bending forces in covalent bonds as nearest and next nearest 
neighbor distance constraints. This simple global counting of constraints
is commonly referred to as Maxwell counting, which may result in positive
or negative values for $F$. A negative $F$ indicates the network is 
over-constrained. Philips \cite{Philips} qualitatively explained why
covalent glass networks with low average coordination form more easily. 
Shortly afterward, the notion of rigidity percolation was introduced by 
Thorpe \cite{Thorpe0}, where depending on chemical composition a network
would microscopically be in a floppy or rigid state, having a well defined
rigidity percolation threshold. Experiments \cite{Angell,Boolchand} 
have shown many physical properties in covalent glasses are related to 
the rigidity transition. In spite of the unique insight that the theory 
of network rigidity offers, it is unfortunate that it still remains a
relatively obscure subject. An authoritative source on concepts of 
rigidity and its broad range of interdisciplinary  applications can be 
found in Ref. \cite{RT_book}.

Network rigidity exhibits long-range character \cite{Guyon} that makes 
calculating properties difficult using brute force methods on elastic 
networks \cite{Feng}. However, the mathematics of first order graph 
rigidity \cite{Crapo,Tay,Graver} 
referred to in the physics literature as {\em generic rigidity} greatly 
simplifies calculations \cite{Jacobs0,Moukarzel0}. Atomic coordinates
are not required in generic rigidity. Only the connectivity property 
of the network is important, making it possible to calculate many 
static mechanical properties exactly using an integer based combinatoric
algorithm. In particular, the exact number of internal independent
degrees of freedom can be calculated, all rigid substructures can be
identified as well as all correlated motions that couple the network of 
rigid clusters. One such algorithm, referred to as the {\em pebble game}, 
is available for general networks in two dimensions \cite{Jacobs1} and 
for bond-bending networks in three dimensions \cite{Jacobs2}. A
bond-bending network has the property that all angles between the 
central-force constraints that stem outward from an atom are fixed. 
In addition, dihedral angles can be constrained. 

Covalent glasses are ideal systems to model as a quenched bond-bending 
network, where there is a natural separation between hard-strong forces 
(central and bond-bending forces) and soft-weak forces (torsional and 
non-bonding forces). The large gap in force strength justifies the 
treatment of covalent glass networks at room temperature to be modeled 
as a mechanical network --- essentially a $T=0$ calculation. Recently, 
constraint counting has been applied to protein structure \cite{Jacobs3}
where covalent bonds, salt-bridges, hydrogen bonds and torsional forces 
on resonant bonds (the peptide bond, for example) were modeled as
mechanical distance constraints. By treating the folded protein structure
as a quenched mechanical bond-bending network, flexible and rigid regions 
were identified and found to correlate well with biologically relevant 
motions. Network rigidity in proteins has also been found to correlate 
with protein folding pathways \cite{Rader,Hespenheide}. The success of
the $T=0$ calculations on protein structure suggest that the folded 
state of the protein acts very much like a mechanical machine under
the conditions responsible for the native fold to be
thermodynamically stable. This result is reassuring, as it
has been well appreciated that protein function is very precise in
its response to molecules it encounters having a high degree of 
specificity that makes it appear to respond like a mechanical machine. 
This empirical observation motivated the use
of network rigidity calculations at $T=0$ in the first place. 
In spite of the success that many mechanical aspects of a protein 
fold can be quantitatively characterized, it is also well 
known \cite{Dill,Rose} that protein stability is a result of a
delicate balance between many weak non-covalent interactions. 
In particular, enthalpic and entropic contributions must be part
of the ledger of accounts to understand protein stability.

The study of protein stability has motivated this work in generalizing
the concept of network rigidity to be applicable at finite temperatures
in physical systems having interactions that do not divide into strong 
and weak compared to $kT$. When viewing a protein as a mechanical network,
two serious problems immediately become apparent. Firstly, hydrogen bonds
are continually breaking and forming consistent with thermal fluctuations, 
and secondly, hydrogen bonds have a wide variety of strength that is
dependent on their local environment \cite{Habermann, H-bond}. In prior
work an energetic cut-off criterion \cite{Jacobs3,Jacobs4} was introduced
to determine a set of hydrogen bonds to model as a constraint. As the
energy cut-off was varied, a hierarchical analysis of rigid clusters was
used to characterize the protein structure. Unfortunately, the energy
cut-off was not directly related to thermodynamic stability, nor the 
entropy from molecular flexibility was considered, which limited the
range of validity of the (T=0) rigidity model to be near the native
structure. These problems can be resolved by modeling microscopic
interactions as distance constraints, where each distance constraint
represents a free energy component within the system. Assigning free
energy contributions to specific types of interactions is commonly done
to interpret experimental measurements and in theoretical discussions
on protein stability \cite{fed1,fed2}. However, the utility of such  
a decomposition is questionable because in general it is not possible 
to obtain the total free energy by simply summing the free energy 
components \cite{Mark}. It will be shown that the free energy of a
system can be obtained from its free energy components by employing 
network rigidity calculations at finite temperature, which combines
mechanical and thermodynamic points of views. 

In section \ref{sec:DCM} A Distance Constraint Model (DCM) is introduced
that enables the partition function to be calculated in terms of an 
ensemble of mechanical frameworks. After the concept of a constraint 
is generalized to contain thermodynamic information, each mechanical
framework of constraints provides an underlying interaction that 
couples enthalpic and entropic terms appearing in Boltzmann factors.
In section \ref{sec:toy} two simple 2-dimensional toy models are worked
out to illustrate the details involved in a calculation. As a final
example, an exact solution of a distance constraint model for homogeneous
peptides that undergo an alpha-helix to coil transition is considered
in section \ref{sec:helix-coil}. In section \ref{sec:discussion}, the 
results from all three models are discussed, and the standard Lifson-Roig 
model for a helix-coil transition is compared with the DCM. Conclusions
are made in section \ref{sec:conclusion}.

\section{Distance Constraint Model\label{sec:DCM}}

Lord Kelvin said, "I never satisfy myself until I can make a mechanical
model of a thing. If I can make a mechanical model I can understand it!"
The distance constraint model (DCM) that will be introduced and carefully
discussed in the following subsections closely adheres to Kelvin's belief.
The objective is to use a mechanical model to understand thermal
stability in biopolymers (the focus of this paper) as well as other
systems such as formation of chalcogenide glasses.

The DCM begins by representing a macromolecule and interactions therein
as a mechanical bar-joint framework. For a single static structure,
generic network rigidity properties can be calculated exactly using a
graph-algorithm that does not depend on geometrical coordinates of atoms,
but only on the {\em topological} arrangement of distance constraints.
Network rigidity is used here as an umbrella-phrase to refer to the
following mechanical properties of a bar-joint framework:

\begin{enumerate}
\item{    Identification of all rigid clusters, where each distinct cluster of
atoms forms a rigid body.}
\item{ Identification of all over-constrained regions, within which elastic
strain energy resides.}
\item{Identification of all flexible regions, wherein the atomic structure
can continuously deform.}
\item{ Identification of all independent constraints and degrees of
  freedom.}
\end{enumerate}

These basic mechanical properties are quite useful in characterizing a
single static structure. In this paper, we will generalize the
mechanical description (at T=0) by employing an ensemble-based approach
to account for thermodynamics. Thermodynamics determines the fate of a
biopolymer, albeit kinetic detours and traps. For example, a protein
unfolds when an increase in conformational entropy outweighs a gain in
enthalpy from an associated loss of many favorable intra-molecular
noncovalent interactions. Furthermore, a functional protein in the
native state is stable against thermal fluctuations through
enthalpy-entropy compensation.

The distance constraint model (DCM) uses network rigidity as an
underlying interaction. Through non-local mechanical interactions,
network rigidity answers the question about which degrees of freedom are
independent and directly relates to the non-additively of measured
component free energies. Although the total enthalpy is additive, the
entropy is not. This non-additive property of component entropies
derives from not knowing which degrees of freedom in the system are
independent or redundant. However, generic network rigidity properties can 
be calculated exactly with the pebble game by recursively adding one 
constraint at a time to build a framework. As constraints are added, some
atoms will become part of a rigid cluster. A new constraint is redundant
when added to an already rigid region and independent when it removes a degree
of freedom. All distance constraints are treated the same in the pebble
game, and there is a clear distinction between a constraint and degree
of freedom.

In the DCM, interactions are represented as distance constraints, each
characterized by an enthalpy and an entropy contribution assumed to
depend only on local structural properties. Constraints are quantified
as being strong or weak based on their entropy contribution. A (greater,
lesser) entropy contribution implies a (weaker, stronger) constraint.
The key aspect of the DCM is that stronger constraints must be placed in
the network before weaker ones in order to generalize network rigidity
to finite temperatures. This leads to a preferential ordering, which is
implemented operationally as:

\begin{enumerate}
\item{Sort all constraints based on entropy assignments in increasing 
order, thereby ranking them from strongest to weakest.}
\item{Add constraints recursively one at a time using the pebble game
according to the rank ordering from strongest to weakest, until the
entire structure is completely rigid.}
\end{enumerate}

The DCM is mathematically well defined and physically intuitive. The
essential idea is that weak constraints allow more conformational
freedom than do strong constraints. Stronger constraints take precedence
in defining rigid structures because weaker constraints are more
accommodating. Thus, a weak constraint acts like a degree of freedom
relative to a strong constraint. Consequently, {\em the notion of a 
constraint and degree of freedom cannot be distinguished clearly
once entropy price tags are introduced}. Rather, a quantitative
measure for conformational entropy is obtained for the structure, 
whereas the $T=0$ style of constraint counting simply regards the
structure as completely rigid. In this way the DCM provides a natural
mechanism for enthalpy-entropy compensation. For example, if by some
fluctuation a strong constraint breaks (such as a hydrogen bond) there
will be a destabilizing gain in enthalpy, but also a compensating gain
in conformational entropy as a weaker constraint substitutes. 
The technical aspects and mathematical details of the DCM are now 
addressed.

\subsection{Relating Thermodynamics to Constraint Topology}

The distance constraint model (DCM) views a physical system at a
coarse-grain level as defining a mechanical bar-joint 
framework. A framework is constructed from distance constraints
that are used to represent microscopic interactions. Each distance
constraint defines an equation of the form $R = constant$,
where $R$ is the distance between a pair of atoms. A microscopic
interaction involving a group of atoms (more than two) can be 
modeled by more than one distance constraint, where the collection 
of distance constraints between different pairs of atoms are
simply referred to as a {\em constraint} (without the word
{\em distance} as a qualifier). A hydrogen bond is an example of
a many body interaction that will be modeled as a particular type
of constraint consisting of three (pairwise) distance constraints.
The enthalpy and entropy contributions from a specific type of 
interaction characterize the corresponding constraint type.
Therefore, let ($\Delta H_t,\Delta S_t$) be the change in 
(enthalpy, entropy) that quantifies constraint type, $t$, when it 
is added to a framework. Over the ensemble of all accessible atomic
configurations, the many different geometries between atoms will 
potentially result in a vast number of constraint types that must
be introduced. However, as demonstrated below, a remarkably few
number of constraint types will often be sufficient to 
quantitatively capture the essential physics. 

The microstates of a system are specified in terms of mechanical
frameworks, ${\cal F}$, where each framework uniquely defines 
the {\em topology of all distance constraints}. The DCM is built 
upon the idea that each framework, ${\cal F}$, having a specific 
topology represents a mini-ensemble of bar-joint networks of 
strict distance constraints within the tolerance allowed by the
geometrical coarse graining. One framework consists of many 
possible atomic-coordinate realizations of strict distance 
constraints. However, because generic rigidity properties are sought
that do not depend on the geometrical details of atomic coordinates,
each realization in this mini-ensemble has exactly the same 
network rigidity properties. Thus, the framework label, ${\cal F}$, 
represents an ensemble of bar-joint frameworks sharing identical
network rigidity properties that are calculated using strict 
distance constraints. 

The relation to thermodynamics can be made because a framework uniquely
identifies a mini-ensemble having constant constraint topology,
enabling a free energy, given as $G( {\cal F} )$, to be meaningfully
assigned. To this end, the total enthalpy of a framework is given by
\begin{equation}
\Delta H( {\cal F} )  = \sum_{t} \Delta H_t N_t( {\cal F} )
\label{eq:DH}
\end{equation}
where $N_t$ is the number of {\em constraints} of type $t$ that are 
present. By exploring all accessible {\em atomic configurations}, an 
ensemble of frameworks (each representing a distinct topology) is
generated. The ensemble of frameworks partitions phase space into 
discrete parts, each having a constant enthalpy over a limited range
of conformational freedom. Therefore, the partition function is given by 
\begin{equation}
Z = \sum_{\cal F}   \Omega( {\cal F} ) e^{ -\beta \Delta H( \cal F ) } 
\label{eq:PF1}
\end{equation}
where $\Omega ({\cal F} )$ is the conformational degeneracy of
framework ${\cal F}$. 

The novel aspect of the DCM is that the conformational entropy, given by
$\Delta S( {\cal F} ) = k \ln \Omega( {\cal F} )$, is obtained by
adding component entropies over independent distance constraints that
are explicitly identified using generic rigidity. Simply adding component
entropies over {\em all} distance constraints will generally lead to a
drastic overestimate for $\Omega ( {\cal F} )$. However, identification 
of whether a distance constraint is independent or redundant is not
unique. The freedom in choosing which distance constraints are
independent is akin to the freedom in choosing an independent basis 
set of vectors to span a vector space. Consequently, the addition 
of component entropies over independent distance constraints will lead
to multiple answers for $\Delta S( {\cal F} )$ if based on arbitrary 
selections. Therefore, an auxiliary {\em preferential selection
criterion} for how to choose the optimal set of independent distance
constraints is required. The crucial part of the DCM is that it
enforces a preferential selection criterion that corresponds to 
the determination of the {\em minimum} possible value of 
$\Delta S( {\cal F} )$.

The total conformational entropy for framework, ${\cal F}$, is 
given by  
\begin{equation}
\Delta S( {\cal F} )  = \sum_{t} \Delta S_t I_t^{(p)}( {\cal F} )
\label{eq:DS}
\end{equation}
where $I_t^{(p)}$ is the number of independent {\em distance 
constraints} of type $t$ present in the framework as determined 
by the preferential, $^{(p)}$, selection criterion. The method
for determining linearly independent constraint equations 
involves building a basis set by iteration, where a new constraint
equation is checked for independence against the current basis 
set. If the new constraint equation is independent, then the
basis set expands. The procedure is continued until all distance
constraints in the framework are checked. The preferential
selection criterion is defined as: {\em Distance constraints 
with lower component entropies take precedence in the order 
that they are checked for linear independence}. By applying 
the preferential selection criterion in conjunction with exact 
constraint counting for generic rigidity, the change in Gibbs
free energy for framework, ${\cal F}$, is given as
\begin{equation}
\Delta G({\cal F}) = \Delta H({\cal F}) - T \; \Delta S({\cal F})
\; \geq \sum_t \Delta G_t({\cal F}) \;\;.
\label{eq:GFE}
\end{equation}
Only in the case that {\em all} distance constraints in the
framework are independent will $\Delta G({\cal F})$ be equal 
to a straightforward sum over the component free energies,
$\Delta G_t({\cal F})$, associated with each constraint type.
The partition function is calculated as 
\begin{equation}
Z = \sum_{\cal F}  e^{ -\beta \Delta G( \cal F ) } 
\label{eq:PF2}
\end{equation}
in accordance to the standard form, except each microstate
corresponds to a generic mechanical framework, ${\cal F}$,
made up of (infinitely strong) holonomic distance constraints, 
and the ensemble consist of all topologically distinct frameworks.

\subsection{Generic Rigidity and Non-Additivity of Entropy}

Meaningful thermodynamic properties are directly tied to local
atomic structure because of coarse graining over geometrical 
bins. To reflect the geometrical aspect of the DCM, the index 
$t$ is represented by two indices $(i,q)$, where $i$ now 
specifies the type of constraint and $q$ labels a specific 
geometrical bin. For example, a hydrogen bond is a particular
type of interaction, but its strength depends on its local
geometry. The component free energy of the $i$-th type of
microscopic interaction is expressed as a free energy 
function, $\Delta G_q^i$, which accounts for all atomic
positions of the group of atoms under consideration within 
the $q$-th geometrical bin. The process of obtaining a 
free energy decomposition \cite{Mark} (the set of $\Delta G_q^i$ 
used in the model) is not unique because different types of
interactions will involve one or more overlapping atoms. Also
there will be unavoidable many body effects, such as electrostatic
interactions between the atoms of interest with all other atoms,
including those in solvent. The non-uniqueness of a free energy  
decomposition can be used as an advantage in the process of 
defining constituent types of constraints.

An effective strategy in employing the DCM is to define a minimum 
number of constraint types with limited number of geometries that 
will yield a desired level of accuracy in predictions. For each
$(i,q)$, the enthalpy and entropy contributions denoted as 
($\Delta H_q^i$, $\Delta S_q^i$) can in principle be determined
self-consistently in lieu of not being unique. Self consistency is 
satisfied when the free energy assignment to small clusters of atoms
used in defining constraint types {\em locally} obey the preferential
selection criterion. This means that various clusters of atoms (for
example, those within an amino acid or hydrogen bond) define subsystems
that are treated the same way as the full system. Knowing the 
thermodynamic properties of a cluster of atoms allows constraint
types to be defined and characterized with a $\Delta H_q^i$ and
$\Delta S_q^i$. It is worth mentioning that in principle, a 
{\em hierarchical} set of constraint types can be constructed 
iteratively by defining constraints with lowest component entropies
first, and in succession defining constraints with the next
lowest component entropy, etc. 

The procedure to determine the local thermodynamic functions
($\Delta H_q^i$, $\Delta S_q^i$) for all constraint types and 
their geometries constitutes a preliminary step in the DCM. In
principle explicit calculations for $\Delta G_q^i$ could be 
made using accurate physical theories (i.e. quantum mechanical
calculations) involving clusters of atoms within a coherent
potential approximation scheme. This type of 
bottom up approach should be tractable and the results would
be very useful. However, these difficult calculations
can be sidestepped (completely or in part) by writing down 
the parametric form of a microscopic free energy function with
empirically derived parameters. The important outcomes are: 
1) interactions are modeled as constraints characterized by
two quantities ($\Delta H_q^i$, $\Delta S_q^i$) that can 
be determined by theoretical means or fitting to large sets
of experimental data, and 2) the DCM parameters can be 
expected to be transferable between systems that are well
described by the same set of constraint types.

The DCM invokes a probabilistic interpretation that all 
distance constraint realizations between atoms are uniformly 
distributed within a geometrical bin. By allowing each atom a 
finite amount of freedom, it is ensured that the framework can 
be treated as generic. Although there will be configurations 
that have atypical atomic positions, these will be of zero
measure. Therefore, the system is modeled as a collection of
generically placed holonomic constraints, for which many  
mechanical properties can be calculated using exact constraint 
counting algorithms. The connection to thermodynamics enters 
into the rigidity calculation by determining the correct 
Boltzmann weight assignment to each mechanical framework,
which is related to the non-additivity of component entropies.
The selected set of independent constraints under the
preferential selection criterion does not depend on 
coordinates in so far that the same framework is maintained
over a {\em limited range of conformational freedom}. This
limited range of conformational freedom is quantified by the 
total entropy, $\Delta S( {\cal F} )$, which depends strongly
on the topology of distance constraints present in the system. 

Calculating the exact value for $\Delta S( {\cal F} )$ will
unfortunately not be possible in the DCM. The preferential 
selection criterion is enforced to obtain the best estimate for 
each framework. Fundamentally, overlap in phase space can occur 
when two constraints are independent but not orthogonal. The
direct result of adding component entropies associated with only
independent constraints is that less phase space will be ``double 
counted''. Therefore, adding component entropies over independent
constraints gives an {\em upper bound} for $\Delta S( {\cal F} )$. 
The preferential selection criterion ensures the lowest possible 
upper bound because the {\em strongest} distance constraints
defined by the smallest entropies are taken as 
independent before {\em weaker} distance constraints having 
larger entropies. The utility of the DCM will depend on the
degree of accuracy in estimating conformational degeneracy. 
Note that distance constraints not sharing atoms are orthogonal,
and do not contribute in over-counting phase space. Although the
distance constraints that share atoms will not generally be
orthogonal, by construction of a self-consistent hierarchical 
series of constraints, phase space overlap between themselves
locally is correctly taken into account. Therefore an accurate 
$\Omega ({\cal F})$ can be expected by using a {\em complete
set} of self consistent constraint types. The phrase ``complete
set'' is used to mean that for any position of atomic coordinates,
{\em a framework is always defined such that after all constraints
are placed it is rigid}. As more constraint types are defined, a
framework becomes increasingly more over constrained, which can 
only lead to a better lowest upper bound.

The preferential selection criterion has a simple physical 
interpretation. Each constraint that is added to a system can 
potentially reduce entropy. However, a redundant distance 
constraint does not reduce entropy \cite{footnote1}. This is
because when a constraint is added to a rigid region that is
formed by stronger constraints, the weaker constraint will
accommodate the structural geometry dictated by the cohort 
of stronger constraints \cite{footnote2}. The strength of a
constraint (strong or weak) is tied to phase space volume.
Therefore, a clear distinction between a constraint and degree 
of freedom is not possible. The rigidity calculation at finite
temperature treats constraints and degrees of freedom on equal
footing in the sense that weaker constraints act as degrees of 
freedom relative to stronger constraints. The entropy loss 
associated with an over constrained region is paid at a 
premium by the strongest member constraints. Fortunately, the
{\em pebble game} algorithm \cite{Jacobs1, Jacobs2} for
determining distance constraint independence is based on a
recursive procedure of building a framework one constraint at a
time. The new implementation only requires using a presorted list
of distance constraints from strongest to weakest. It is worth 
noting that this algorithm does not model a kinetic process as
the constraints in a particular framework are present all the time.
Rather, the entropy loss from a constraint has to do with its
{\em effectiveness} relative to all other constraints in the framework.

\subsection{Quenched and Fluctuating Constraints}

The term {\em quenched constraint} refers to a constraint type
that will be present among a particular group of atoms in all
frameworks of the ensemble. For example, over the temperature
range of biological importance, covalent bonding between atoms
within a protein is modeled as a set of quenched constraints.
Furthermore, the central and bond-bending forces that make up
covalent bonding are modeled by constraints having microscopic
free energies associated with a single geometrical bin. The
torsional force component will also be modeled by a quenched 
constraint (as the torsional force is always present) but
will have a microscopic free energy, $\Delta G_q^i$, with 
multiple geometrical bins (labeled by $q$) depending on the
dihedral angle. A system modeled by a complete set of 
quenched constraints will generally be associated with an 
ensemble of frameworks because the enthalpic and entropic
characteristics of distance constraints depend on local
geometry. In the extreme case where only one framework is
accessible, the DCM will not provide optimal accuracy whereas
normal mode analysis is more appropriate. For example, if
a FCC solid is modeled using one central force constraint
type, then the DCM is equivalent to the Einstein model.

The term {\em fluctuating constraint} refers to a constraint 
type that may or may not be present among a particular group
of atoms having a fixed geometry. When a fluctuating constraint
is present, it is associated with a microscopic free energy,
$\Delta G_q^i$, in the same way as a quenched constraint. However,
a fluctuating constraint is not strictly tied to geometry
because it may not be present. The DCM allows for fluctuating
constraints to account for degrees of freedom (dof) that are 
not explicitly part of a system. For example, solvent dof couple
to protein atoms defining a system. The solvent-protein 
interactions are modeled as fluctuating constraints on the 
system. In this way, hydration shells around protein atoms are 
modeled as fluctuating constraint types characterized by enthalpy, 
$\Delta H_q^i$, and entropy, $\Delta S_q^i$, contributions 
that account for the many body interactions. Even more basic
is the hydrogen bond. Hydrogen bonding is modeled as a
fluctuating constraint because 1) the protein atom electronic
dof are not explicitly described, and 2) solvent dof compete 
with intra-molecular hydrogen bonding for a given geometry. 
Thus, the DCM provides a real-space description involving
mechanical constraints that directly accounts for fluctuating 
hydrogen bonding, such as found in proteins and water. 

\subsection{Temperature Independent Model Parameters}

The enthalpy and entropy contributions, ($\Delta H_q^i$, 
$\Delta S_q^i$) assigned to the various constraint types are
functions of temperature, pressure, and other thermodynamic 
conditions dealing with the chemical environment, such as
pH, ionic strength or whether the local geometry is in a
hydrophobic or polar neighborhood. Therefore, caution must be
exercised in the ordering of the constraints from strongest
to weakest, because this ordering may change
as the thermodynamic conditions change. Consequently, the
environmentally induced re-ordering of constraint types
by relative strength could potentially cause dramatic
conformational change. However, the utility of the DCM can
best be appreciated by using a simplified description.

\begin{figure}
\begin{center}
\epsfysize=9cm\epsfbox{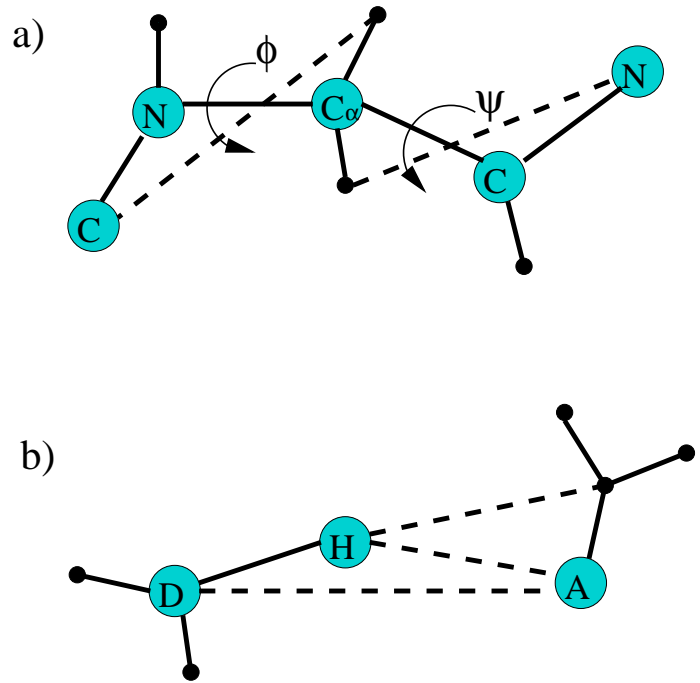}
\caption{
     (a) The torsion interaction within the backbone of an amino acid
         is modeled by two distance constraints shown as dashed lines
         that lock two dihedral angles. Except for proline, the
         torsion constraint is characterized by $(V_q, 2 \delta_q )$,
         where $q$ denotes a geometrical bin within a two dimensional
         ($\phi,\psi$)-space. When the geometry is such that both
         angles fall within region $q$, the energy is $V_q$ and each
         distance constraint has $\delta_q$ pure entropy. For the
         $\alpha$-helix to coil transition, $q$ represents either a
         $\alpha$-helical or coil geometry.
     (b) The hydrogen bond interaction is modeled by three distance
         constraints shown as dashed lines. The hydrogen bond
         constraint is characterized by $(U_q, 3 \gamma_q )$, where
         $q$ labels geometrical bins that can be defined in different
         ways. For the $\alpha$-helix to coil transition, $q$
         represents the geometry formed by spanning across three
         consecutive amino acids that can independently be in either
         $\alpha$-helical or coil geometries.
        }
\label{fig:hierarchical2}
\end{center}
\end{figure}

Model parameters will be taken as constants. Furthermore, the
entropic term will be distributed equally over all the distance
constraints that model a particular constraint type. Then, all
microscopic free energies will have the generic form
\begin{equation}
\Delta G_q^i = \epsilon_q^i - T k \; m^i \gamma_q^i
\label{eq:MGFE}
\end{equation}
where $\epsilon_q^i$ is energy,
$\gamma_q^i$ is a dimensionless variable referred to as pure
entropy, and $m^i$ is the number of distance constraints that
are used to model the $i$-th constraint type. Pure entropies
are taken to be positive because they are fundamentally related to 
the number of accessible quantum states that are associated with 
a specified geometrical bin tolerance, given by 
$e^{\gamma_q^i} \geq 1$.
Figure \ref{fig:hierarchical2} shows two example constraint 
types that will be used in section \ref{sec:helix-coil} to model
an $\alpha$-helix to coil transition. A constraint type is now
generically characterized as $(\epsilon_q^i , m^i \gamma_q^i)$. 
These parameters can be interpreted as being derived by Taylor
expanding to first order the true microscopic Gibbs free energy 
about some temperature of most interest. Analogous to
Eq. (\ref{eq:DH}), the total energy of a framework is given as
\begin{equation}
E( {\cal F} )  = \sum_{ \{ (i,q)_j \} } \epsilon_q^i 
                 { \eta_q^i}_j  ( {\cal F} )
\label{eq:total_E}
\end{equation}
where ${\eta_q^i}_j ( {\cal F} )$ equals (1,0) when the $j$-th 
constraint of the $i$-th type is (present, or not present) within
the $q$-th geometrical bin. Analogous to Eq. (\ref{eq:DS}) the total 
pure entropy of a framework is given as 
\begin{equation}
\tau ( {\cal F} )  = \sum_{ \{ (i,q)_j \} } \gamma_q^i 
                     {\sigma_q^i}_j ( {\cal F} )
\label{eq:total_PE}
\end{equation}
where ${\sigma_q^i}_j ( {\cal F} )$ is the number of independent 
distance constraints within the $j$-th constraint of the system,
in accordance with generic rigidity and the preferential selection 
criterion. Note that $\{ 0$, $1$, $\dots m^i \}$ are the possible 
values that ${\sigma_q^i}_j ( {\cal F} )$ can take.

The partition function is now written as 
\begin{equation}
Z = \sum_{ {\cal F} } e^{\tau ( {\cal F} ) } e^{ - \beta E( {\cal F} ) } 
\; = \;
\sum_{ {\cal F} } \prod_{ \{ (i,q)_j \}_{ {\cal F} } } 
                  e^{ \gamma_q^i {\sigma_q^i}_j } \;
                  e^{ -\beta \epsilon_q^i {\eta_q^i}_j }
\label{eq:PF3}
\end{equation}
where the form of Eq. (\ref{eq:PF3}) suggests that the energy and 
entropy contributions are independent. However, not only do the
values of \{ ${\sigma_q^i}_j$ \} depend on calculations from 
generic rigidity, but when ${\eta_q^i}_j ({\cal F})=0$ then 
${\sigma_q^i}_j ({\cal F})=0$. Thus, the energy and entropy of each
framework is coupled through topology via the underlying interaction
of network rigidity. For example, consider the entropy loss 
associated with the formation of a hydrogen bond. As shown in
Fig. \ref{fig:hierarchical2}b the hydrogen bond constraint
is modeled as three distance constraints. For a particular
geometry, the hydrogen bond contributes energy $U_q$ and it
contributes $\{ 0$, $\gamma_q$, $2\gamma_q$, $3\gamma_q \}$
amount of pure entropy to the system, depending on if it has
$\{ 0$, $1$, $2$, $3 \}$ independent distance constraints. If
$\gamma_q$ is comparatively small indicating a {\em relatively}
strong distance constraint, then the greatest entropy loss 
for the system occurs when all three distance constraints are
independent. In contrast, if $\gamma_q$ is comparatively large
indicating a {\em relatively} weak distance constraint, then 
the greatest entropy loss for the system occurs when all three
distance constraints are redundant. Since the results depend
on the topological arrangement of all constraints
in the system, {\em no a priori statement can be made about 
whether the formation of a hydrogen bond will supply a 
favorable or unfavorable entropic contribution}. 

\section{Toy Models in Two Dimensions\label{sec:toy}}
\begin{figure}
\begin{center}
\epsfysize=6cm\epsfbox{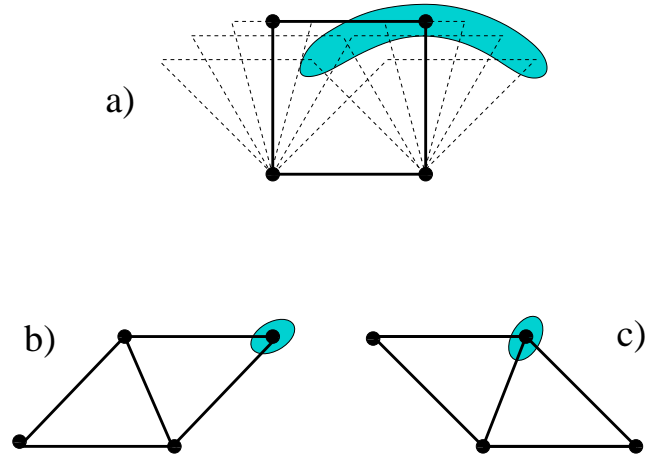}
\caption{
          A small two dimensional ring molecule in the shape of a
          quadrilateral. The shaded regions schematically show the
          allowed geomentrical variation for fixed topology indictive
          of the degree of flexibility. Configuration (a) is topologically
          distinct from (b) and (c). For identical atoms at each corner,
          configurations (b) and (c) represent the same topology of
          constraints, but are distinct otherwise. The framework in (a)
          is refered to as state H; where it has greater energy and
          conformational entropy than (b/c). The (b and/or c) framework
          is refered to as state L; where the bond along the diagonal
          leads to a lower energy and conformational entropy than (a).
        }
\label{fig:square}
\end{center}
\end{figure}

The (internal) partition function for the two dimensional molecule
shown in Fig. \ref{fig:square} is calculated to illustrate basic
concepts. The molecule consists of four identical atoms
connected together by four strong central force bonds forming
a quadrilateral. The central force (cf) bonds are modeled as
quenched constraints characterized by energy, $U_{cf}$ and pure
entropy $\gamma_{cf}$. Four torsional forces are also modeled
as quenched constraints. In 2D the torsion force (tf) is
a function of the angle between a pair of central force bonds.
It is modeled as a next nearest neighbor distance constraint
characterized by energy $V_{tf}$ and pure entropy $\delta_{tf}$.
The torsional free energy surface is assumed shallow over a
large range of angles. 
A hydrogen bond (hb) in 2D is considered a single central force,
and is modeled as a fluctuating constraint characterized by
energy $U_{hb}$ and pure entropy $\gamma_{hb}$. Within a
length tolerance, a hydrogen bond can form between a pair of
atoms along either diagonal of the quadrilateral.

As Fig. \ref{fig:square} shows, there are only two distinct types 
of frameworks, labeled as $(L$ and $H)$ when the hydrogen bond
(is, is not) present. This is a two level system (three states 
are required for distinguishable atoms). Employing the DCM,
the first step is to rank order the distance constraints from
strongest to weakest. This ranking is based on sorting the
pure entropies from lowest to highest, assumed given as:
\begin{equation}
\begin{array}{rccccc}
\mbox{pure entropy:} & \gamma_{cf} & < & \gamma_{hb} & < & \delta_{tf} \\
        \mbox{rank:} &      1      &   &    2        &   &     3
\end{array}
\label{sq:pe_ordering}
\end{equation}
The second step requires calculating the total energy and pure
entropy for each framework using the preferential selection
criterion. In state $H$ there are 8 distance constraints
(4 cf and 4 tf) and in state $L$ there are 9 distance constraints
(4 cf, 4 tf and 1 hb). There are 8 dof, three of which involve
global translations and rotations. Five distance constraints will
always be independent making the framework rigid. From
Eqs. (\ref{eq:total_E} and \ref{eq:total_PE}) it follows that:
\begin{equation}
\begin{array}{rll}
\mbox{State H:} &  \tau_H = 4 \gamma_{cf} + \delta_{tf}    &
            \;\;\;\;\;  E_H = 4 U_{cf} + 4 V_{tf}         \\
\mbox{State L:} &  \tau_L = 4 \gamma_{cf} + \gamma_{hb}    &
            \;\;\;\;\;  E_L = 4 U_{cf} + 4 V_{tf} + U_{hb}
\end{array}
\label{sq:Z_terms}
\end{equation}
Therefore, the (internal) partition function is given as
\begin{equation}
Z = e^{\tau_L} e^{ -\beta E_L } + e^{\tau_H} e^{ -\beta E_H } \;\; .
\label{eq:2level0}
\end{equation}

With $U_{hb} < 0$ as expected for chemical bonding, the states
$(L$, $H)$ will be more probable at (low, high) temperatures
respectively. Since for both states, the energy and pure entropy
terms associated with the cf-constraints and the energy terms for
the tf-constraints are the same, the partition function simplifies to
\begin{equation}
Z = Z_o \; [ e^{ \gamma_{hb} } e^{ -\beta U_{hb} } + e^{ \delta_{tf} } ]
\label{eq:2level}
\end{equation}
where $Z_o$ contains the terms common in both $L$ and $H$ states.
This example illustrates a general result that the strongest quenched
constraints play a passive role. Molecular cooperativity is controlled
by competition among weaker interactions. It is worth mentioning that
if the two level approximation does not produce a sufficiently accurate
temperature response, then the model parameters could be regarded as
temperature dependent functions. Alternatively, the single geometrical
bin for the assumed weakly varying (as a function of temperature)
torsional free energy can be further subdivided to better account for
thermodynamic response by creating more terms in the partition function.

\begin{figure}
\begin{center}
\epsfysize=10cm\epsfbox{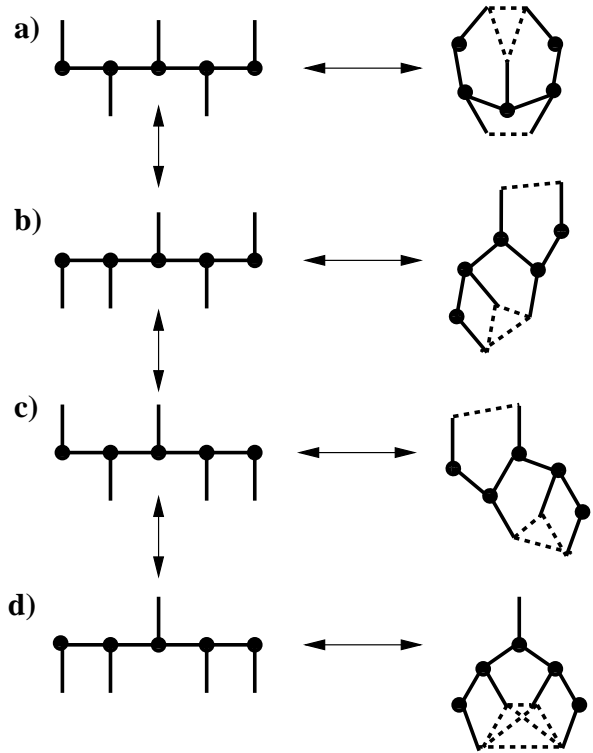}
\caption{ A small two dimensional chain molecule. Backbone atoms are
          denoted by filled circles. There are only four topologically
          distinct conformations (a-d) accessible to the molecule as it
          deforms during the process of hydrogen bonds breaking and
          reforming.  Dashed lines represent
          hydrogen bonding. Left side: All conformations have a
          large conformational degeneracy. Right side: When all
          hydrogen bonds are present the molecule has much less
          conformational degeneracy. In particular, for conformation
          (d) a Rigid state is defined when all four side-chain atoms
          form a rigid cluster from hydrogen bonding.
        }
\label{fig:mol}
\end{center}
\end{figure}

The (internal) partition function for a more interesting two
dimensional molecule shown in Fig. \ref{fig:mol} is calculated.
This molecule consists of five backbone and five side-chain atoms
connected by central forces. A side-chain atom at the end of the
chain can swing around the backbone atom, but it is assumed
that a potential barrier must be crossed. The highest point
of the energy barrier is when the side-chain atom is co-linear
with the backbone chain. Therefore, the molecule is regarded to
have four topologically distinct conformations, each having the
same characteristic energy basin. Finally, side-chain atoms that
are in sufficient proximity of one another can hydrogen bond.

The central force constraint is characterized by ($U_{cf}$,
$\gamma_{cf}$), and the hydrogen bond constraint is characterized
by $(U$, $\gamma)$. There are two types of torsion
force constraints involving angles between $BBB$ atoms or
$BBS$ atoms, where $B$ and $S$ represent backbone and side-chain
atoms respectively. The torsional constraint type for the
$BBB$ angle is characterized by $(V_{BBB}$, $\delta_{BBB})$ and
the torsional constraint type for the $BBS$ angle is
characterized by $(V$, $\delta)$. The distance constraints
are now ranked from strongest to weakest, assumed given as
\begin{equation}
\begin{array}{rccccccc}
\mbox{pure entropy:} & \gamma_{cf} & < & \gamma & < &
                                \delta  & < & \delta_{BBB} \\
        \mbox{rank:} &      1      &   &    2        &   &
                                    3   &   &      4
\end{array}
\label{mol:pe_ordering}
\end{equation}
Since both torsion constraint types are quenched constraints, it
follows that the pure entropy parameter for the $BBB$ type of
angle is always irrelevant for all frameworks in the ensemble.
This example illustrates an important point that weak forces
often need not be associated with an entropy term, because
they will always be redundant. Nevertheless, many weak forces can
still play an important role in the energetics.

There are a total of 112 possible frameworks, corresponding to
$2^4$ different frameworks (due to fluctuating hydrogen bonds)
for each of the topologically distinct conformations
shown in Fig. \ref{fig:mol}abc and $2^6$ frameworks for the
conformation shown in Fig. \ref{fig:mol}d. Once all the
central force constraints are placed (first) there are 8 internal
dof remaining in the molecule. If no hydrogen bond constraints
are placed, then the total pure entropy of the molecule will
be $9 \gamma_{cf} + 8 \delta$, which gives the maximum
possible value. As hydrogen bond constraints are added, the
total pure entropy will decrease. The best chance of finding
a redundant hydrogen bond is when the maximum number is 
present for each distinct topology. By inspection, only one 
framework out of 112 has a redundant hydrogen bond constraint,
corresponding to the six hydrogen bonds all simultaneously
present in the conformation shown in Fig. \ref{fig:mol}d. 
Recall that the parameters associated with the quenched
constraints common to all frameworks can be factored out.
Therefore, relative to the conformations containing no
hydrogen bonds, the change in Gibbs free energy, 
$\Delta G(n)$, for the molecule having $n$ hydrogen bonds
is given by
\[
\left\{
\begin{array}
[c]{lll}
\Delta G(n) = n \; U - k T [n \; \gamma + (8-n) \; \delta] & 
                                \mbox{for} & n = 0, 1, \dots 5; \\
\Delta G(6) = 6 \; U - k T [5 \; \gamma +   3   \; \delta] &  &
\end{array}
\right.
\]
The factor of $(8-n)$ appears because each independent hydrogen bond 
constraint eliminates an angular dof. The remaining (weakest) 
torsion force constraints rigidifies the molecule.

In this example many of the frameworks have degenerate 
Gibbs free energy. The Gibbs free energy already accounts 
for conformational degeneracy, but there is also a 
configurational degeneracy in the number of hydrogen
bond combinations that are possible. Therefore, the partition
function is written as
\begin{equation}
Z = \sum_{n} g(n) \; e^{ - \beta \Delta G(n) }
\end{equation}
where $g(n)$ is the number of frameworks with $n$ hydrogen
bonds. The values of $g(n)$ for different $n$ are tabulated
in Table \ref{table:gn}, which is obtained by straightforward
counting.

\begin{table}[t]
\begin{tabular}
[c]{c|ccccccc}\hline \hline
$n$    & 0 & 1  & 2  & 3  & 4  & 5 & 6 \\\hline
$g(n)$ & 4 & 18 & 33 & 32 & 18 & 6 & 1 \\
\hline \hline
\end{tabular}
\caption{Hydrogen bond configurational degeneracy.
\label{table:gn}
        }
\end{table}

\begin{figure}
\begin{center}
\epsfysize=7cm\epsfbox{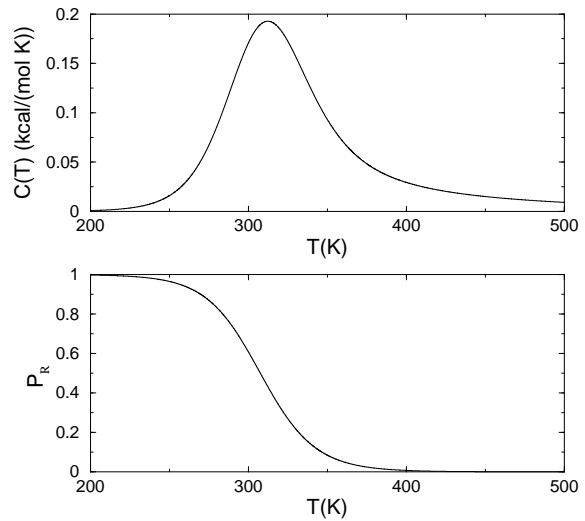}
\caption{
         (a) Heat capacity as a function of temperature.
         (b) Probability for the molecule to form a rigid
             structural unit. The selected parameters were
             obtained by choosing the marked point on the phase
             diagram in Fig. \ref{fig:phase}, and fixing the
             transition temperature to be near 310 Kelvin.
        }
\label{fig:C_p}
\end{center}
\end{figure}

\begin{figure}
\begin{center}
\epsfysize=7cm\epsfbox{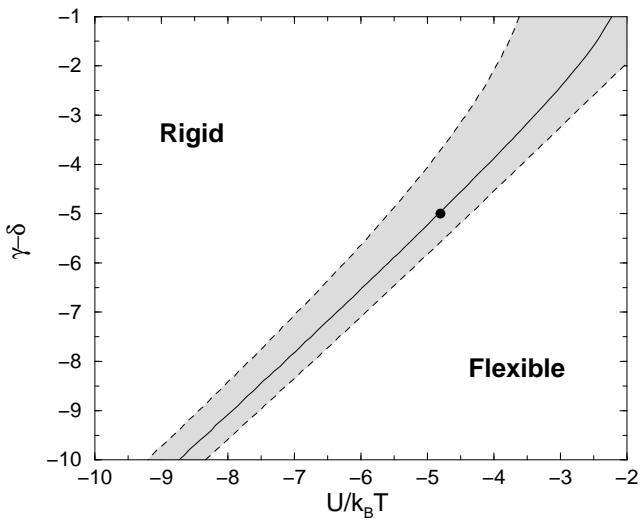}
\caption{The phase diagram of the two dimensional chain molecule.
         The difference in pure entropy between the hydrogen bond
         and torsional force constraints, and the
         hydrogen bond energy scaled by thermal energy are the
         only two relevant variables.
        }
\label{fig:phase}
\end{center}
\end{figure}

The heat capacity is plotted in Fig.\ref{fig:C_p}a, showing a
peak near 310 Kelvin, where the model parameters were fixed to
convenient values to show interesting features. This peak is a
manifestation of a structural transition from the {\em Rigid}
state (defined in Fig. \ref{fig:mol}d) at low temperature to a
{\em Flexible} state at high temperature. The degree of rigidity
is also shown by plotting the equilibrium probability, $P_R$, for
the molecule to be described by a framework with 5 or 6 hydrogen
bonds, where
\begin{equation}
P_R(T) = ( e^{ -\beta \Delta G(6) } + g(5) \; e^{ -\beta \Delta G(5) } )/Z
\label{eq:P_R}
\end{equation}
represents only the frameworks that form a rigid structural unit.
The probability for being in the rigid state is used as an order
parameter. A phase diagram is shown in Fig. \ref{fig:phase}, where
the solid line corresponds to the maximum heat capacity used to locate
the transition temperature. The shaded area defines a broad transition
region defined as $(0.1 < P_R < 0.9)$ indicating no substantial
preference for either the rigid or flexible states.

\section{Alpha Helix to Coil Transition\label{sec:helix-coil}}

\begin{figure}
\begin{center}
\epsfysize=6cm\epsfbox{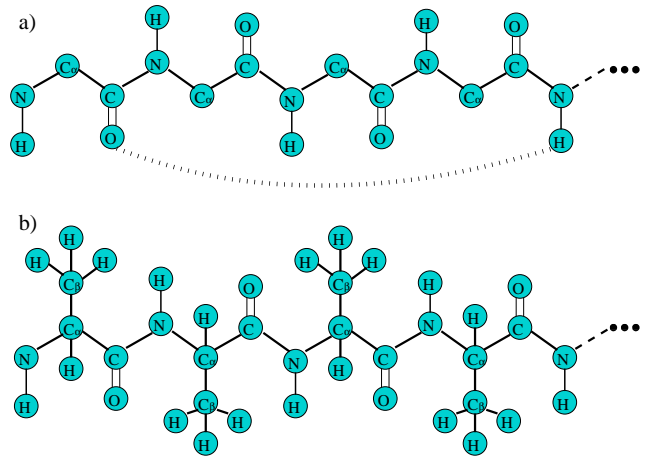}
\caption{(a) The backbone of a peptide chain. The dihedral angle
             of the peptide bond (C-N) cannot rotate. The long
             curved dashed line represents a possible hydrogen bond.
         (b) An example of poly-alanine. The dihedral angle
             between C$_\alpha$-C$_\beta$ can rotate.
        }
\label{fig:peptidechain}
\end{center}
\end{figure}

The DCM is employed to describe a transition from a stable
$\alpha$-helix structure that is rigid at low temperature to a
flexible coil involving many disordered conformations at high
temperature. The backbone of a homogeneous peptide chain, as depicted
in Fig. \ref{fig:peptidechain}a, is considered for simplicity. Compared
to the Zimm-Bragg \cite{Zimm} or Lifson-Roig \cite{Lifson} models,
the DCM is mathematically more complicated because network rigidity is
a long-range interaction that will be explicitly quantified in terms
of a direct product between a rigidity state space and a conformational
state space, from which a transfer matrix is constructed.

Four constraint types are used here to model central, bond-bending
and torsional forces involved in covalent bonds as well as hydrogen
bonds. The strongest two constraint types, modeling the central and
bond-bending forces, are placed in the network before the weaker
constraint types. Thus, a chain of $n$ amino acids has $2n$ dof
along the backbone because
only the $\phi$ and $\psi$ dihedral angles in each amino acid
(proline is is not considered here) are free to rotate. The energy
and pure entropy parameters for the central and bond-bending
constraint types are not of concern because they play a passive role
in the partition function, as explained in section \ref{sec:toy}. The
remaining two constraint types depend on the local conformation of
the backbone as determined by the $\phi$ and $\psi$ dihedral angles.
Explicit side-chain to side-chain and side-chain to backbone
interactions are not considered in the analysis given here.

The third constraint type describes a torsion interaction. Torsion
constraints along the backbone are partitioned into distinct
geometrical bins depending on the $\phi$ and $\psi$ angles. For
example, different bins can be defined using a Ramachandran plot
\cite{Ramachandran,R-plot} for each type of amino acid. Here,
the $\alpha$-helical and coil geometries, labeled $a$ and $c$
respectively, are considered to be the only two accessible
conformational states. The coil geometry, $c$, includes all
other secondary structures (non $\alpha$-helical) such as a
$\beta$-strand, 3-10 helix or left-handed $\alpha$-helix. The
(energy, pure entropy) of the $\alpha$-helical and coil torsion
constraints are given by $(V_a$, $2\delta_a)$ and $(V_c$,
$2\delta_c)$ respectively. As shown in Fig. \ref{fig:hierarchical2}a,
the torsion constraint contains two distance constraints to lock
the $\phi$ and $\psi$ angles. Each distance constraint carries
a pure entropy of $\delta_a$ or $\delta_c$ in the $\alpha$-helix
or coil geometry respectively.

The fourth constraint type describes hydrogen bonding. For
simplicity, only backbone hydrogen bonds between the carbonyl
oxygen of the $i$-th amino acid and the amine nitrogen of the
$(i+4)$-th amino acid are considered accessible. The (energy,
pure entropy) for a hydrogen bond constraint are given by
($U_{xyz}$, $3\gamma_{xyz}$) where $x$, $y$ and $z$ specify the
local ($a$ or $c$) backbone geometries of the $i+1$, $i+2$ and
$i+3$ amino acids that are spanned. As shown in 
Fig. \ref{fig:hierarchical2}b, a hydrogen bond constraint contains 
three distance constraints, where each distance constraint carries
a pure entropy of $\gamma_{xyz}$. Noting that there are eight 
possible geometries, each requiring the two parameters $U_{xyz}$
and $\gamma_{xyz}$, gives a tally of 16 parameters for the
hydrogen bond constraint type.

The peptide chain is decomposed into triplets, denoted by $[xyz]_i$, 
where $x$, $y$ and $z$ represent $a$ or $c$ geometries for the 
$\{ i$, $i+1$, $i+2 \}$ amino acids. To account for hydrogen bond 
fluctuations, a triplet may or may not have a spanning hydrogen bond.
Another variable, $\lambda_i = (1,0)$ is used to specify whether a
hydrogen bond constraint is (present or not) across the $i$-th 
triplet. When present, a hydrogen bond spans the $i$-th triplet by 
connecting the $i-1$ amino acid to the $i+3$ amino acid. The greatest 
number of hydrogen bonds that can form within an $\alpha$-helix of $n$
amino acids is $n-4$, since the only triplets that can have a spanning
hydrogen bond are $(i=2, \; 3, \;  \dots , \;  n-3)$. Note that the 
variable $\lambda_i$ corresponds to the $i$-th amino acid in the chain,
and therefore it is associated with the leading edge of a triplet. A
triplet (not at the ends) will have 16 possible conformational states 
corresponding to 8 different local-geometries with or without a 
hydrogen bond. The complete specification of the conformation of
a triplet has the general form $\lambda [xyz]$. An energy $U_o$ 
is introduced for triplets of the form $0[xyz]$, which represents
the hydrogen bond energy resulting between the peptide backbone
and solvent. Therefore, $U_o$ is an additional hydrogen bond
parameter (17 total) in the DCM considered here. 

\subsection{Rigidity Propagation Rule}

To facilitate exact constraint counting subjected to the preferential
selection criterion, the degree of rigidity for a triplet is 
specified by a {\em local rigidity state}, denoted as 
$| \mbox{lrs} \rangle$.
The local rigidity state contains the minimum amount of information
about rigidity at the end of a chain such that when the next amino
acid is added, the local rigidity state of the end triplet is uniquely 
specified. The set of all accessible local rigidity states, 
$\{ | \mbox{lrs} \rangle \}$, will serve as a basis set for a rigidity
state space. A complete basis set will be generated using the rigidity
propagation rule. 

Each triplet has 6 dof, 6 torsional force distance constraints and 
when there is a spanning hydrogen bond 3 additional hydrogen bond
distance constraints. The pure entropies of each type of distance 
constraint is rank-ordered from 1 to 10 because there are 8 different
$\gamma_{xyz}$ and 2 different $\delta_x$ assuming no degeneracies. 
A torsional force distance constraint (tfdc) and a hydrogen bond
distance constraint (hbdc) lock dihedral angles differently. A tfdc 
is confined to lock a specific dihedral angle, whereas a hbdc spans
all 6 dof within a triplet. A hbdc can be used to lock any of these
6 dof, and should lock the one which will minimize the total 
conformational entropy of the chain. In this sense, hydrogen bond 
distance constraints are promiscuous. Consequently, the dof that is
best to lock cannot be determined solely on the local triplet conformation
because network rigidity is a long-range interaction. Therefore, an 
algorithm for propagating the local rigidity state must be established. 

A local rigidity state specifies the {\em current} 
rank assignment of constraints used to lock the first 4 dof in a 
triplet. The rank assignment corresponds only to {\em independent}
constraints. The local rigidity state is represented as
\begin{equation}
| \mbox{lrs} \rangle = | r_1, r_2, r_3, r_4 \rangle
\label{eq:lrs_template}
\end{equation}
where $r_k$ is the rank of the distance constraint that locks the
$k$-th dihedral angle in a triplet. The ranks of the last two
dihedral angles within a triplet will become important in determining
the local rigidity state of the next triplet upon propagation.  
The explicit form for $| \mbox{lrs} \rangle$ in Eq. (\ref{eq:lrs_template})
provides a book-keeping device to calculate the preferential sum
of pure entropies over independent constraints. The algorithm for
propagating rigidity from left to right takes the form:

\begin{enumerate}

\item Given $| r_1, r_2, r_3, r_4 \rangle$: 
      Retain the 4 {\em temporary} rank assignments and augment the
      2 ranks from the torsional constraint on the third amino acid,
      thus forming a temporary template involving 6 ranks, given by:
      $\{ r_1, r_2, r_3, r_4, r_5, r_6 \}$

\item If no hydrogen bond is present continue to the next step.
      Otherwise, perform the following operations when a hydrogen 
      bond spans the new triplet. Attempt to place one distance
      constraint at a time, each having a rank of $r_{hb}$. Find
      the maximum rank, denoted as $r^{(1)}_{i}$, out of the six 
      current ranks in the template. The superscript $^{(1)}$
      indicates that this is the maximum rank, and the index $i$ 
      specifies its location within the template. If 
      $r_{hb} \geq r^{(1)}_i$, continue to the next step because 
      this and any of the remaining hydrogen bond distance constraints 
      are redundant. Otherwise, replace the maximum rank by $r_{hb}$.
      Working from right to left (the direction opposite to
      propagation) find the next maximum rank, denoted as 
      $r^{(2)}_{j}$. If $r^{(2)}_{j} > r_{hb}$ then swap ranks.
      That is, let $r_i = r^{(2)}_{j}$ and $r_j = r_{hb}$. 
      Continue the process of swapping rank $r_{hb}$ with the next
      greatest rank to its left, until it can no longer be shifted
      to the left. Continue to the next step when all three 
      hydrogen bond distance constraints have been place.

\item The first two degrees of freedom in the triplet are 
      permanently locked by distance constraints that are 
      associated with the ranks $r_1$ and $r_2$ in the template. The
      remaining four ranks in the template define the current
      local rigidity state of the new triplet given as:
      $|r_1^{\prime} = r_3,$  $r_2^{\prime} = r_4,$ 
       $r_3^{\prime} = r_5,$  $r_4^{\prime} = r_6 \rangle$.
      Repeat this process (back to step 1) until the propagation
      through all triplets is finished.

\end{enumerate}

Step 2 can be understood conceptually. Ranks within a
template act as a dof relative to a hbdc rank whenever they
are greater than $r_{hb}$, otherwise they act as a constraint.
Among the ranks acting as a dof, a lower rank acts as a
constraint relative to a greater rank. Therefore, the
greatest rank should be replaced by $r_{hb}$. However, it
could happen in a future test (as the chain is propagated 
from left to right) that the largest rank within the current 
template could be replaced by a different hbdc that spans
a different triplet downstream. If this happens, it would 
be better to use the current hbdc to lock the second highest
rank. Replacing the highest rank, or replacing the second 
highest rank, etc, depends on the relative rank of a future 
hbdc, if any appear at all! This makes the transfer matrix
approach different than the usual case, because rigidity is
non-local where the conformations down the chain will affect 
the optimal rank substitution at the current triplet. 

The first hbdc encountered down the line that overlaps with
part of the current triplet will be effective as a constraint
{\em within the current triplet} only if its rank is lower than
the greatest rank, $r^{(1)}$ found in Eq. (\ref{eq:lrs_template}). 
The second effective hbdc must have a rank lower than the 
second greatest rank, $r^{(2)}$. If no effective hbdc is 
encountered, it is best to replace $r^{(1)}$ with $r_{hb}$ 
in step 2 of the algorithm. If one effective hbdc is encountered, 
it is best to replace $r^{(2)}$ with $r_{hb}$. More generally, if
$n$ effective hbdc are encountered, it is best to replace
$r^{(n+1)}$ with $r_{hb}$ if possible. All these cases are
properly handled by building into the definition of a local
rigidity state a {\em chain reaction} that automatically swaps
higher ranks into lower ranks {\em when needed}. The chain
reaction is initialized in step 2 by the process of swapping 
ranks within a triplet from highest to lowest working in the 
{\em opposite} direction of propagation. The outcome of the
above algorithm, is that both the long-range interaction of 
rigidity and the global preferential selection criterion are
properly described.

\begin{figure*}
\begin{center}
\epsfysize=12cm\epsfbox{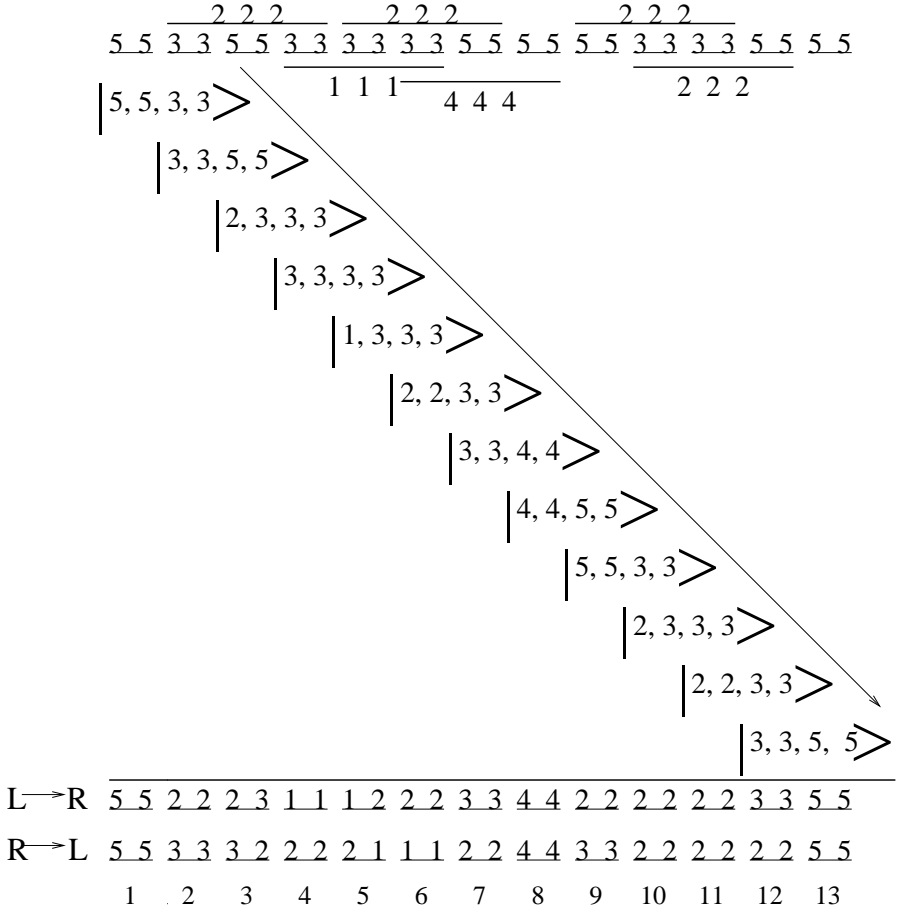}
\caption{
        The top schematic describes the backbone of a 13-mer peptide chain in
        a conformation that has: Torsional constraints with pure entropy
        ranked either 3 or 5, and occasional hydrogen bond constraints
        (pictorially represented as bars spanning three consecutive pairs of
        dihedral angle dof) with pure entropy ranked either 1, 2 or 4. Each
        step in the propagation of the local rigidity state from left to
        right is shown along the diagonal. The final ranks that remain after
        propagating from left to right (L $\rightarrow$ R) is given on the
        3rd to last row. The final ranks obtained by propagating from right
        to left (R $\rightarrow$ L) is given on the 2nd to last row. The
        last row labels the amino acids. Both propagation directions yield
        3 rank-1, 12 rank-2, 5 rank-3, 2 rank-4 and 4 rank-5 independent
        distance constraints.
        }
\label{fig:propagate}
\end{center}
\end{figure*}

Figure \ref{fig:propagate} shows how the rigidity propagation
rule is implemented on a short chain in a particular framework.
The initial description of the chain includes the ranks of all
torsion and hydrogen bond constraints that are present. This
framework contains 18 redundant constraints since the chain in
{\em any conformation} is always just rigid (isostatic) whenever
there are no hydrogen bonds along the backbone, and here there
are 3 $\times$ (6 hydrogen bonds) extra distance constraints.
The final description shows the ranks of only independent distance
constraints that remain after being permanently assigned in step 3
of the propagation rule. The final ordering of ranks generally
depends on the direction of propagation, but the final distribution
of ranks (i.e. number of independent constraints having rank
1, 2, $\dots$) is invariant. Moreover, the final rank-distribution
is identical to that of a preferential selected set of independent
constraints obtained by placing the strongest distance constraints
before weaker ones in otherwise arbitrary order.

Referring to Fig. \ref{fig:propagate}, the entire process of
propagating from left to right is shown. The 1st triplet has a local
rigidity state given by $|5, 5, 3, 3 \rangle$. This 1st triplet
does not have a spanning hydrogen bond, therefore, the next triplet
(after the first propagation) has a local rigidity
state given by $|3, 3, 5, 5 \rangle$. During the first propagation
step, each tfdc within the 1st amino acid is recorded as independent,
locking the $\phi_1$ and $\psi_1$ dihedral angles. The pure entropy
associated with these two distance constraints are recorded in terms
of the two ranks, $\{5,5\}$. For the second propagation step, the
spanning hydrogen bond across the 2nd triplet changes the temporary
rank assignments as follows:
\begin{equation}
\begin{array}{rcccccc}
\mbox{initial $| \mbox{lrs} \rangle$:} &  |3, & 3,  & 5, & 5 \rangle &     &     \\
\hline
   \mbox{tfdc template:} & \{3, & 3,  & 5,  & 5,        &    3,     & 3\} \\
   \mbox{plus 1st hbdc:} &      &     &     & 2,        &           &     \\
  \mbox{intermediate 1:} & \{2, & 3,  & 3,  & 5,        &    3,     & 3\} \\
   \mbox{plus 2nd hbdc:} &      &     &     & 2,        &           &     \\
  \mbox{intermediate 2:} & \{2, & 2,  & 3,  & 3,        &    3,     & 3\} \\
   \mbox{plus 3rd hbdc:} &      &     &     &           &           & 2\;\;\; \\
  \mbox{intermediate 3:} & \{2, & 2,  & 2,  & 3,        &    3,     & 3\} \\
\hline
\mbox{final $| \mbox{lrs} \rangle$:} &  &     & |2, & 3, & 3,    & 3 \rangle      \\
\end{array}
\label{ex:rigidity_calc1}
\end{equation}
\vskip1cm
The $\phi_2$ and $\psi_2$ are considered to be locked by two of the
promiscuous hydrogen bond distance constraints, and recorded by the
two ranks $\{2,2\}$.

The rigidity propagation rule applied to a specified framework,
${\cal F}$, allows the total pure entropy, $\tau ( {\cal F} )$,
to be calculated as the sum over pure entropies associated with the
ranks of the distance constraints used to permanently lock the
$\phi$ and $\psi$ dof. For a given framework, the alternative
calculation for $\tau ( {\cal F} )$ is to use the {\em pebble
game} algorithm \cite{Jacobs1, Jacobs2}, where the distance
constraints with lowest ranks are placed in the network first.
The propagation algorithm was explicitly tested \cite{prop_test}
against exact calculations using the {\em pebble game}. Although
preferential constraint counting offers an exact calculation method,
by incorporating the rigidity propagation rule into a transfer
matrix, $\tau ( {\cal F} )$ no longer requires {\em explicit}
calculation on each framework in the ensemble.

\subsection{Transfer Matrix and the Partition Function}

The transfer matrix is constructed from a direct product space
formed by a triplet conformational state denoted by
$| \lambda, x,y,z \rangle$, where $\lambda$ is one when a hydrogen bond
spans the $x,y,z$ triplet, zero otherwise and $x, y, z$ are either
alpha helix (a) or coil (c). A triplet is completely specified as
\begin{equation}
\mbox{triplet state} = | \lambda, x,y,z \rangle
                   \otimes | r_1, r_2, r_3, r_4 \rangle \;\;\; ,
\label{eq:triplet_state}
\end{equation}
where $r_1$ and $r_2$ are the ranks of the constraints on the phi and psi angle (backbone angles) of
the $x$ state, and $r_3$ and $r_4$ are the corresponding ranks of the
constraints on the $y$ state.  The 4 ranks on the first two amino acids,
the presence or absence of a
spanning hydrogen bond, and the conformational state (helix or coil)
of each residue together completely specify a state.

Most elements of the transfer matrix, $T$, will be zero. The non-zero
matrix elements have the form given by:
\begin{eqnarray}
\langle \lambda^{\prime}, x^{\prime} = y, y^{\prime} = z, z^{\prime} |
        \otimes   \langle r^{\prime}_1, r^{\prime}_2,
                r^{\prime}_3, r^{\prime}_4  | T
 | \lambda, x, y, z \rangle \nonumber \\
  \otimes | r_1, r_2, r_3, r_4 \rangle
       = e^{ \Delta \tau_p } e^{ - \beta \; \Delta \epsilon_p }
\label{eq:matrix_element}
\end{eqnarray}
where after a propagation to the right the new first amino acid
corresponds to the prior middle amino acid and the new middle amino acid
corresponds to the prior right amino acid. In addition to this, the
matrix element will only be non-zero if the set of final ranks in
the local rigidity state obey the rigidity propagation rules. The
non-zero matrix element then contributes a Boltzmann factor that accounts
for both the energy and pure entropy contributions of the constraints
encountered. The variables $\Delta \tau_p$ and $\Delta \epsilon_p$
respectively represent the change in pure entropy and energy upon
propagation along the chain. The contribution to $\Delta \tau_p$ at
each propagation step is given by the sum of pure entropies of the
two constraints that permanently lock the two dof within the first
amino acid of a triplet. Thus $\Delta \tau_p$ is determined by
the rigidity state space in accordance to step 3 of the rigidity
propagation rule. In contrast, $\Delta \epsilon_p$ is determined
by the conformational state space where it is a function of only
$\lambda [xyz]$ and it is found by summing the hydrogen bond energy
given by $U_{xyz}$ when $\lambda =1$ and $U_o$ when $\lambda =0$,
with the torsional force constraint energy given by $V_x$.
By construction, the zeros and non-zeros of the transfer matrix
accounts for the rigidity propagation rules, thereby correctly
propagating rigidity.

Ignoring boundary conditions momentarily, the (internal) partition function
could be calculated as:
\begin{equation}
Z_n = \langle f | T^{n} \; | i \rangle
\label{eq:Z_peptide_noEdge}
\end{equation}

The method for constructing the
transfer matrix, $T$, is explained by working through an example.
Consider a chain of 13 amino acids where the framework given as
\begin{equation}
\begin{array}{ccccccccccccccc}
{\bf 0} & 1 & 0 & 1 & 1 & 1 & 0 & 0 & 1 & 1 & {\bf 0} & {\bf 0} & {\bf 0} &   &   \\
     c  & a & c & a & a & a & c & c & c & a & a & c & c & s & s
\end{array}
\label{ex:framework}
\end{equation}
is one realization taken from an ensemble of $2^{(13+9)}$ frameworks
describing all accessible chain conformations (of a chain of length 13).
The numbers of 1 or 0 on
top of an $a$ or $c$ specify $\lambda$ in a triplet, $\lambda[xyz]$.
A number placed over an amino acid describes a hydrogen
bond that spans it and the next two amino acids to the right. In
order for a chain of length $n$ to be represented by $n$-triplets, two
$s$ solvent states are explicitly shown as being augmented at the right
end of the chain. Effects of this state are discussed below under
boundary conditions.  The first and last three zeros (in bold)
correspond to triplets for which an intra-molecular hydrogen bond cannot form.

The dimension and form of the transfer matrix, $T$, strongly depends on
the rank ordering of pure entropies. For purpose of illustration,
consider the rank ordering:
\begin{equation}
\begin{array}{rccccccccccccc}

\mbox{pure entropy:} & 0 & < & \gamma_{aaa} & < &
         \left\{
            \begin{array}{l}
            \gamma_{caa} \\ \gamma_{aca} \\ \gamma_{aac}
            \end{array}
         \right.
                         & < &   \delta_a   & < & 
         \left\{
            \begin{array}{l}
            \gamma_{cca} \\ \gamma_{cac} \\ \gamma_{acc}
            \end{array}
         \right.
                         & < &   \delta_c   & < & \gamma_{ccc} \\ 

        \mbox{rank:} & 0 & & 1 &  & 2 & & 3 & & 4 & & 5 & & 6
\end{array}
\label{ex:pev_ordering}
\end{equation}
where rank 0 is associated with the special $s$-conformation, and rank
6 is associated with a hydrogen bond that spans a local $[ccc]$ geometry.
In this case, $\gamma_{ccc}$ plays no role because it will always be
redundant. In this example, intra-molecular hydrogen bonds that span the same
number of coil states within a triplet are degenerate. Thus,
$\gamma_{caa} = $ $\gamma_{aca} = $
$\gamma_{aac}$ and $U_{caa} = $ $U_{aca} = $ $U_{aac}$, etc.

The initial product vector that needs to be propagated is given as
$|0, c, a, c \rangle | 5, 5, 3, 3 \rangle$, where the symbol $\otimes$
will be dropped from now on. This vector is obtained below by considering
the process of propagating triplet $0[ssc]$ to $0[sca]$ before arriving
to the current triplet $0[cac]$. Using the rigidity propagation rule,
the 1st matrix multiplication by $T$ propagates the initial vector
into vector $|1, a, c, a \rangle | 3, 3, 5, 5 \rangle$, while the 2nd
matrix multiplication gives $|0, c, a, a \rangle | 2, 3, 3, 3 \rangle$.
The shifts in the conformational states are obvious, and the
propagation of the local rigidity states are calculated according
to Example \ref{ex:rigidity_calc1}. In fact, the initial configuration
of ranks shown in Fig. \ref{fig:propagate} precisely correspond to
the framework given in Ex. \ref{ex:framework}. In the 1st propagation
step, the contribution of pure entropies from constraints that lock
the $\phi_1$ and $\psi_1$ dihedral angles is given as
$\Delta \tau_1 = 2 \delta_c$. The energy contribution is 
$\Delta \epsilon_1 = V_c + U_o$, which reflects the hydrogen bond
energy between peptide and solvent. At each propagation step 
another product vector will be generated. The second step takes the
vector $|1, a, c, a \rangle | 3, 3, 5, 5 \rangle$ into vector  
$|0, c, a, a \rangle | 2, 3, 3, 3 \rangle$. The energy contribution
is $\Delta \epsilon_2 =$ $V_a + $ $U_{aca}$, which reflects the 
intra-molecular hydrogen bond energy that depends on local geometry $[aca]$.
The pure entropy contribution is given by $\Delta \tau_2 =$
$2 \gamma_{aca}$, resulting from two rank 2 pure entropy values. All
matrix elements are  determined by energy contributions from consecutive
triplet conformation states described in Ex. \ref{ex:framework}, and pure
entropy contributions are determined by the final rank ordering (from
left to right) listed in Fig. \ref{fig:propagate}. Some matrix elements
generated by the framework given in Ex. \ref{ex:framework} are listed in
Table \ref{table:T}.

\subsubsection{Boundary Conditions}

In addition to constructing the transfer matrix, $T$, the boundary
conditions on both the left and right ends of the chain must be
specified. The boundary conditions are of particular importance
for peptides that are experimentally studied because most often
they are less than 20 amino acids long. The approach taken here
is to add auxiliary triplet states before and after the chain
to take into account solvation effects. A requirement that the
left and right boundary conditions must satisfy is: left to right
propagation and right to left must yield identical results for all
observable quantities. This basic requirement is satisfied by
the approach used here.

An infinite number of auxiliary $s$-conformations are appended 
to the beginning and end of the chain to represent bulk solvent. 
A triplet of auxiliary $s$-conformations is of the form $0[sss]$,
and it is used as a reference state. The transfer matrix propagates
the triplet $0[sss]$ into another $0[sss]$ triplet with a Boltzmann
weight of 1 {\em by definition}. The auxiliary $s$-conformations play 
a passive role in the calculation (as if they are not present)
except in triplets at the ends of the chain where they mix with
$a$- or $c$-conformations within the chain. Physical boundary
conditions require the local rigidity state of the last $0[sss]$ 
solvent triplet just before the chain to be equal to the local 
rigidity state of the first $0[sss]$ solvent triplet at the end 
of the chain. Furthermore, this local rigidity state must be the 
same for any peptide, regardless of its length or composition. 
Therefore, the local rigidity state for the $0[sss]$ solvent 
triplet is defined as $| r_s, r_s, r_s, r_s \rangle$ where
$r_s \equiv 0$ to represent the lowest rank associated with a
minimum pure entropy, $\gamma_s \equiv 0$, which is the lowest 
physically realizable value. Consequently, when propagating
from one solvent triplet to the next $\Delta \tau_p = 0$, and
by setting $\Delta \epsilon_p \equiv 0$, then the Boltzmann
weight of 1 is ensured. With these boundary conditions no bulk
properties of solvent (the reservoir) are calculated, while
peptide to solvent interactions are taken into account by
fluctuating constraints acting on the peptide (the system).

\begin{table}[t]
\begin{tabular}{|l|l|l|}
\hline
\hline

step & transfer matrix element & Boltzmann factor \\
\hline
 1  & $\langle 1,a,c,a| \langle 3,3,5,5|T|0,c,a,c \rangle |5,5,3,3 \rangle$ &  $e^{2 \delta_c}              e^{ - \beta ( V_c + U_o     ) }$ \\
 2  & $\langle 0,c,a,a| \langle 2,3,3,3|T|1,a,c,a \rangle |3,3,5,5 \rangle$ &  $e^{2 \gamma_{aca}}          e^{ - \beta ( V_a + U_{aca} ) }$ \\
 3  & $\langle 1,a,a,a| \langle 3,3,3,3|T|0,c,a,a \rangle |2,3,3,3 \rangle$ &  $e^{\gamma_{caa} + \delta_a} e^{ - \beta ( V_c + U_o     ) }$ \\
 4  & $\langle 1,a,a,c| \langle 1,3,3,3|T|1,a,a,a \rangle |3,3,3,3 \rangle$ &  $e^{2\gamma_{aaa}}           e^{ - \beta ( V_a + U_{aaa} ) }$ \\
\vdots &                       \vdots                                       &        \vdots                                                  \\
10  & $\langle 0,a,c,c| \langle 2,2,3,3|T|1,a,a,c \rangle |2,3,3,3 \rangle$ &  $e^{2 \gamma_{caa}}          e^{ - \beta ( V_a + U_{aac} ) }$ \\
11  & $\langle 0,c,c,s| \langle 3,3,5,5|S|0,a,c,c \rangle |2,2,3,3 \rangle$ &  $e^{2 \gamma_{aac}}          e^{ - \beta ( V_a + U_o     ) }$ \\
12  & $\langle 0,c,s,s| \langle 5,5,0,0|R|0,c,c,s \rangle |3,3,5,5 \rangle$ &  $e^{2 \delta_a}              e^{ - \beta ( V_c + U_o     ) }$ \\
13  & $\langle 0,s,s,s| \langle 0,0,0,0|Q|0,c,s,s \rangle |5,5,0,0 \rangle$ &  $e^{2 \delta_c}              e^{ - \beta ( V_c + U_o     ) }$ \\
\hline
\hline
\end{tabular}

\caption{ A short list of selected matrix elements that are generated
          from the framework given in Example \ref{ex:framework}.
          Refer to Fig. \ref{fig:propagate} to check the
          correspondence between the pure entropy contribution
          $\Delta \tau_p$ on the $p$-th propagation step with
          the final ranks listed for left to right propagation.
        \label{table:T}
        }
\end{table}

Consider propagating from left to right. Then the left boundary 
condition is most conveniently represented as a column vector in 
the direct product space, denoted as $| i \rangle$. The form of 
the initial vector is given by  
\begin{equation}
| i \rangle = \sum_{x, y, z} \; 
  e^{ - \beta ( \Delta \epsilon_{0ssx} + \Delta \epsilon_{0sxy} ) } \;
  | 0, x, y, z \rangle \otimes | r_x, r_x, r_y, r_y \rangle \;\;\; .
\label{eq:initial_vect}
\end{equation}
The ranks $r_{x}$ and $r_{y}$ are respectively associated with the
pure entropy of a tfdc in conformation state ($x$ of the 1st amino
acid) and ($y$ of the 2nd amino acid). No entropic contributions
arise in propagating from the $0[sss]$ triplet to the $0[xyz]$
triplet because of the rigidity propagation rule when no hydrogen
bonds are present and the definition of the special $s$-conformation.
However, $\Delta \epsilon_{0ssx}$ and $\Delta \epsilon_{0sxy}$ account
for solvation energy between the peptide and solvent. Here a 
triplet with no spanning hydrogen bond is taken to contribute
$U_o$ energy. Therefore, the initial state vector simplifies to
\begin{equation}
| i \rangle = \sum_{x, y, z} \;  e^{ - \beta 2 U_o } \;
|0, x, y, z \rangle \otimes | r_x, r_x, r_y, r_y \rangle  \;\;\; .
\label{eq:initial_vect2}
\end{equation}

The right-end boundary condition is implemented using three 
special transfer matrices that involve the $s$-conformation. 
Starting from the $\lambda [xyz]_{n-3}$ triplet, transfer
matrices $S$, $R$ and $Q$ are defined to respectively propagate 
from $\lambda [xyz]_{n-3}$ to $0[yzs]$ to $0[zss]$ and finally
to the $0[sss]$ triplet. These three matrices in succession
channel all possible local rigidity states accessible at triplet
$\lambda [xyz]_{n-3}$ to $| r_s, r_s, r_s, r_s \rangle$ when the 
$0[sss]$ solvent triplet is reached. Therefore, the only
non-zero component in the direct product space after matrix $Q$ 
is applied is given by the vector $|0, s, s, s \rangle$ $\otimes$
$|r_s, r_s, r_s, r_s \rangle$, which is denoted as $| f \rangle$.
By construction, the final vector does not change upon further
propagation from $0[sss]$ to all remaining $0[sss]$ solvent 
triplets \cite{bound_s}. 

Including boundary conditions, the (internal) partition function is
calculated as:
\begin{equation}
Z_n = \langle f | \; Q \; R \; S \; T^{n-3} \; | i \rangle
\; \; \; \forall \; \; \; n \geq 3
\label{eq:Z_peptide}
\end{equation}
for homogeneous peptide chains with $n$ amino acids, and it
involves $n$ matrix multiplications over $n$ triplets. The 
form of Eq. (\ref{eq:Z_peptide}) is independent of the direction
used to propagate rigidity. By inspection the partition 
function for a tripeptide ($n=3$) reduces to
\begin{equation}
Z_3 = e^{ -\beta 5 U_o } \; \sum_{x,y,z} \; 
      e^{ 2( \delta_x + \delta_y + \delta_z ) } \;
      e^{ - \beta ( V_x + V_y + V_z ) } \; \; \; .
\label{eq:tripeptide}
\end{equation}
The expression for $Z_3$ highlights two subtleties about the
simplifying assumptions invoked here that are worth mentioning.
(1) Unlike the intra-molecular hydrogen bonds, the energy, $U_o$, for 
hydrogen bonding between the peptide and solvent is not considered 
to depend on the local peptide geometry (specified by $[xyz]$.) (2)
No pure entropy parameter (given by $\gamma_o$) is associated
with the peptide-solvent hydrogen bonds because it has been
assumed to be larger than all other pure entropies that
characterize the four constraint types introduced above. As illustrated by
the 2nd toy model in section \ref{sec:toy},
constraints having a pure entropy greater than all others that are always redundant
do not contribute entropicly. Not allowing for
entropic contributions from peptide-solvent hydrogen bonds 
implies the solvent molecules (aqueous solution being of primary
interest) are unstructured around the peptide. In other work,
hydration effects due to structured water around the peptide 
is explicitly modeled \cite{Jacobs_DCM2} as an additional
constraint type. 

\subsubsection{Generating the complete basis set}

With Eq. (\ref{eq:Z_peptide}) at hand, what remains is to generate 
the complete basis set of vectors in the product space. This is
done during the process of constructing the transfer matrices. 
The procedure for generating the transfer matrices, $T$, $S$, $R$ and 
$Q$ begins by considering all 8 possibilities for the starting product
space vector. Then propagation to all possible next triplets is
performed. Each distinct vector that is created defines another basis 
vector. For each basis vector that was not previously generated, it
is propagated to all possible next triplets. Eventually the same
vectors continue to be generated by recursively considering all
vectors --- indicating a complete basis set is formed. It is worth
mentioning that the product space is ergodic, in the sense that 
starting from any vector representing a triplet state of the peptide 
chain, any other vector can be reached by some number of transfer 
matrix multiplications. In some cases, this number can be quite
long, depending on the size of the transfer matrix. A priori,
the number of distinct product space vectors is not known because
the number of local rigidity states must be calculated using the
rigidity propagation rule. In Table \ref{table:dim} the dimension,
$M$, of the product vector space is listed for several choices of
rank-orderings. A large matrix size is an indication of the
long-range nature of rigidity that manifests itself as molecular
cooperativity.

\begin{table}[t]
\begin{tabular}
[c]{|c|c|c|c|c|c|c|c|c|c|c|c|c|}
\hline
\hline
               & a  & b      & c      & d  & e  & f  & g   & h   & i   &  j  & k   & l  \\
$\delta_a$     & 1  & 1      & 1      & 2  & 2  & 2  & 2   & 3   & 3   &  4  & 4   & 5  \\
$\delta_c$     & 2  & 3      & 3      & 3  & 4  & 5  & 6   & 5   & 6   &  5  & 6   & 9  \\
$\gamma_{aaa}$ & R  & 1 or 2 & 1 or 2 & 1  & 1  & 1  & 1   & 1   & 1   &  1  & 1   & 1  \\
$\gamma_{caa}$ & R  & R      & 2      & 2  & 3  & 3  & 3   & 2   & 2   &  2  & 2   & 2  \\
$\gamma_{aca}$ & R  & R      & 2      & 2  & 3  & 3  & 3   & 2   & 2   &  2  & 2   & 3  \\
$\gamma_{aac}$ & R  & R      & 2      & 2  & 3  & 3  & 3   & 2   & 2   &  2  & 2   & 4  \\
$\gamma_{cca}$ & R  & R      & R      & R  & R  & 4  & 4   & 4   & 4   &  3  & 3   & 6  \\
$\gamma_{cac}$ & R  & R      & R      & R  & R  & 4  & 4   & 4   & 4   &  3  & 3   & 7  \\
$\gamma_{acc}$ & R  & R      & R      & R  & R  & 4  & 4   & 4   & 4   &  3  & 3   & 8  \\
$\gamma_{ccc}$ & R  & R      & R      & R  & R  & R  & 5   & R   & 5   &  R  & 5   & R  \\
\hline
$Dim(T)$       & 16 & 16     & 28     & 60 & 60 & 96 & 140 & 200 & 244 & 376 & 436 & 444 \\
\hline
\hline
\end{tabular}

\caption{ Some examples of possible results that can be obtained after sorting the
          set of DCM pure entropies from lowest to highest, and then assigned ranks
          of 1 and greater respectively. Pure entropy values listed on the leftmost
          column are rank-ordered in the various columns. The letter R indicates
          that the constraint is always redundant, and therefore is always
          ineffective in reducing conformational entropy. The last row gives the
          dimension \cite{bound_s} of the transfer matrix $T$ for the particular
          rank ordering. Column (h) corresponds to the rank ordering used in Example
          \ref{ex:pev_ordering}. Column (l) is similar to column (h) except that
          degeneracy is lifted from all the $\gamma_{xyz}$ pure entropies.
        \label{table:dim}
        }
\end{table}

\subsection{DCM Results Compared to Monte Carlo Simulation}

The transition from a rigid $\alpha$-helical state to a flexible
coil state is characterized by helix content, which serves as an
order parameter. The helix content is defined as the average fraction
of amino acids in the chain having $\phi$ and $\psi$ dihedral angles
of $\alpha$-helix geometry. The conformational state of the first and
last amino acids are explicitly taken into account. Helix content is
given by the number of amino acids in the $\alpha$-helical
conformation divided by the number of amino acids in the chain.
Applying standard transfer matrix methods, helix content and specific heat
are numerically calculated for any specified set of model parameters. Using 
simulated annealing methods, the DCM parameters were optimized to fit to
Monte Carlo (MC) simulation data \cite{Okamoto} for poly-alanine of 
length 10 in both gas phase (no solvent) and model-water solvent 1,
as well as MC simulation data \cite{Hansmann} for chain lengths of 10, 
15, 20 and 30 in model-water solvent 2. 

The DCM parameters describing the backbone dof for a homogeneous
peptide in solvent include \{ $V_a$, $\delta_a$, $V_c$, $\delta_c$ \}.
Since the amino acids located at the N- and C- termini are exposed to 
solvent differently, it is expected that the backbone parameters 
for the first and last amino acids should be modified. To keep the
number of model parameters to a minimum, the set of parameters given by
\{ $V_a^\prime$, $\delta_a^\prime$, $V_c^\prime$, $\delta_c^\prime$ \}
are used for both the N- and C- termini. Besides these 8 parameters describing
dihedral angle characteristics along the backbone, 17 parameters 
describe hydrogen bonding. To obtain a more manageable number of model 
parameters, many hydrogen bond parameters are considered to be
degenerate, where it is assumed that 1) $U_{cca}$ $=$ $U_{cac}$ $=$
$U_{acc}$, 2) $U_{caa}$ $=$ $U_{aca}$ $=$ $U_{aac}$, 3)  $\gamma_{cca}$ 
$=$ $\gamma_{cac}$ $=$ $\gamma_{acc}$, 4) $\gamma_{caa}$ $=$ 
$\gamma_{aca}$ $=$ $\gamma_{aac}$. This simplification reduces the 
number of hydrogen bond parameters to 9. Taking advantage of the
arbitrariness in absolute energies and entropies, the parameters
$\gamma_{aaa}$, $U_o$, $V_a$ and $V_a^\prime$ can be preset without 
affecting the helix content or the specific heat. Therefore, all
backbone dof are fully described by thirteen $(8 + 9 - 4)$ DCM 
parameters. 

Fitting the DCM to MC simulation of poly-alanine requires
additional parameters to account for the flexibility in the alanine 
side-chain. The side-chain of alanine consists of one dihedral angle
between the C$_\alpha$ and C$_\beta$ atoms as shown in Fig.
\ref{fig:peptidechain}b. An additional torsion constraint type was
applied to this single side-chain dihedral angle. The side-chain torsion
constraint is partitioned into two geometrical bins. Only differences
in energy and pure entropy between the two states are required,
which are characterized by $(V_s$, $\delta_s)$. Since no interactions
are considered between an alanine side-chain with the backbone or other
side-chains, the values of ($V_s$, $\delta_s$) have no affect on helix
content, but do affect specific heat. Another fitting variable, $c_b$, (not
a model parameter) is introduced to represent a constant baseline for the
specific heat. The variable, $c_b$, is required because the DCM is defined
at a coarse grained level, and as such it cannot account for residual energy
fluctuations.

In total, 16 variables are to be determined by fitting to helix content and
specific heat data generated by MC simulation \cite{Okamoto, Hansmann}.
Although each DCM parameter has a physical basis, 16 variables creates
the unfortunate problem that helix content and specific heat can be
simultaneously fitted with a multitude of excellent best fit solutions. This
over parameterization can be quickly avoided, however. An important aspect
of the DCM is that although many parameters have been initially generated
when the set of constraint types were defined for the helix-coil
system; there is no size dependence. Furthermore, the number of parameters
grow slowly when fitting to different solvents because no solvent dependence
is assumed for: 1) intra-molecular hydrogen bond parameters, 2) backbone dihedral
angle parameters not depending on coil conformations, 3) side-chain
dihedral angle parameters and 4) the specific heat baseline.

\begin{figure}
\begin{center}
\epsfysize=7cm\epsfbox{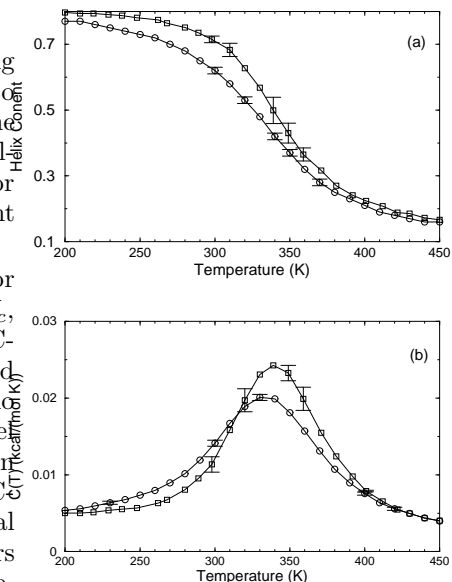}
\caption{
        The (a) helix content and (b) specific heat from two Monte Carlo
        simulations are shown. The deviation between the two simulation
        data for chains of length 10
        creates an intrinsic error that prevents finding a ``good'' fit
        when both results are treated as a single solvent type. Ignoring
        these deviations makes the meaning of goodness not sufficiently
        restrictive, which allows too many ``good'' parameter solutions.
        Instead, this data is treated as two different solvents, where
        (squares, circles) represent model-water 1 and 2 respectively.
        }
\label{fig:discrepancy}
\end{center}
\end{figure}

The cohort of MC data allows 12 curves to be fitted simultaneously.
Superscripts $g$, 1 and 2 are used to respectively refer to gas phase
and model-water solvents 1 and 2. Both model-water solvent 1 and 2 refer
to the MC data generated using the ECEPP/2 force field \cite{ECEPP2}.
Initially, it was assumed that the model-water solvent of both simulations
could be treated identically, since both groups used the same force field.
However, as shown in Fig. \ref{fig:discrepancy} there are sufficient
differences between the chain length 10 data to warrant treating them
as {\em different model-water solvents}. Between the two model-water 
solvents, 10 solvent independent parameters are in common, and (5+5) 
solvent dependent parameters are required. Including the gas phase data 
requires 5 more solvent dependent parameters. In total, 25 fitting
parameters to 12 distinct curves eliminates over-fitting.

\begin{figure}
\begin{center}
\epsfysize=7cm\epsfbox{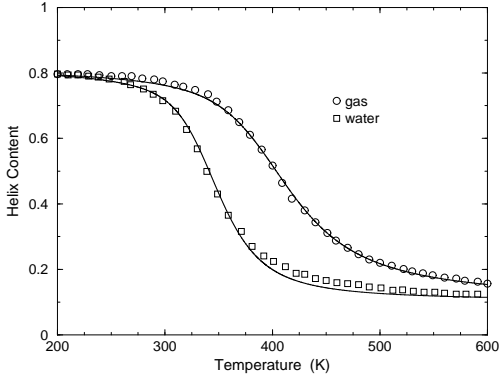}
\caption{
        Best fit to helix content for gas and model-water 1, obtained
        by simultaneous fitting 19 parameters to the cohort of MC data.
        }
\label{fig:Okamoto_opar}
\end{center}
\end{figure}

Interestingly, it was found (from several good best fits) that some
parameters are consistently in close proximity to one another. A
greater fitting error was exchanged for a
maximum reduction of free parameters \cite{best_fit}. Specifically, it
was possible to obtain good fits when forcing different parameters that
were found in close proximity to be equal. This results in demanding:
1) $\delta_c^{1}$ $=$ $\delta_c^{2}$ $=$ $\delta_c^{g}$, 
2) $V_c^1$ $=$ $V_c^g$, 
3) $V_{c}^{\prime ~ 1}$ $=$ $V_{c}^{\prime ~ 2}$,
4) $\delta_{c}^{\prime ~ 1}$ $=$ $\delta_{c}^{\prime ~ 2}$ and 
5) $U_{o}^{2}$ $=$ $U_{o}^{g}$ --- as suggested by the unconstrained fits.
With this reduction, 19 free parameters were used to fit 12 distinct curves
simultaneously.

\begin{table}[b]
\begin{tabular}
[c]{c|c|c|c|c}\hline \hline
               & $aaa$            & $aca$            & $cac$            & $ccc$           \\ \hline
$U_{xyz}$      & -4.637           & -2.827           &  -2.339          & 0.000$^\dagger$ \\
               & -4.95 $\pm$ 0.39 & -3.11 $\pm$ 0.32 & -2.56 $\pm$ 0.33 & 0.000$^\dagger$ \\ \hline
$\gamma_{xyz}$ &  2.000$^\dagger$ &  2.149           &  2.760           & 2.917           \\
               &  2.000$^\dagger$ & 2.19 $\pm$ 0.07  & 2.81 $\pm$ 0.04  & 2.99 $\pm$ 0.12 \\ \hline \hline
\end{tabular}

\vskip 1cm

\begin{tabular}
[c]{c|c|c|c|c|c}\hline \hline
$V_a$           & $\delta_a$       &      $V_a'$     &  $\delta_a'$    &  $V_s$           & $\delta_s$       \\  \hline
0.000$^\dagger$ & 2.656            & 0.000$^\dagger$ & 2.000$^\dagger$ &  1.590           & 3.614            \\
0.000$^\dagger$ & 2.56 $\pm$ 0.24  & 0.000$^\dagger$ & 2.000$^\dagger$ &  1.57 $\pm$ 0.13 & 3.38 $\pm$ 0.15  \\  \hline \hline
\end{tabular}

\caption{Listing 9 solvent independent parameters. Units of energy is in kcal/mol,
         and pure entropies are dimensionless. The numbers with 3 digits
         represent a typical best-fit solution, which were used to generate
         figures \ref{fig:Okamoto_opar}, \ref{fig:Okamoto_c}, \ref{fig:Hansmann_opar},
         \ref{fig:Hansmann_c} and \ref{fig:Tc_lines}. The numbers directly below these
         are the averages obtained over 8 typical simulated annealed fitting solutions,
         including standard deviations for statistical errors. In addition to these
         DCM parameters, the specific heat baseline was considered solvent independent and
         was given as: 0.00133 kcal/(mol K) with average and standard deviation given as
         0.0014 $\pm$ 0.0001 kcal/(mol K). Arbitrarily fixed parameters
         are indicated with a $^\dagger$.
         \label{table:sip}
         }
\end{table}

\begin{figure}
\begin{center}
\epsfysize=7cm\epsfbox{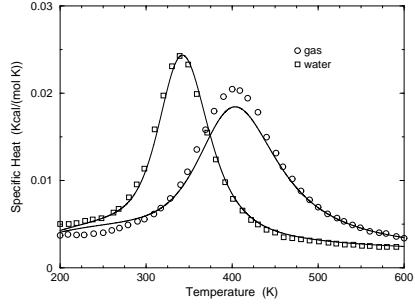}
\caption{
        Best fit to specific heat for gas and model-water 1, obtained by
        simultaneous fitting 19 parameters to the cohort of MC data.
        }
\label{fig:Okamoto_c}
\end{center}
\end{figure}

\begin{figure}
\begin{center}
\epsfysize=7cm\epsfbox{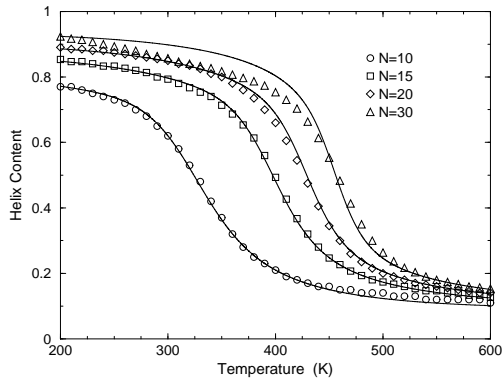}
\caption{
        Best fit to helix content for model-water 2, obtained by
        simultaneous fitting 19 parameters to the cohort of MC data.
        The large deviation seen in the chain of length 30 is at
        an acceptable level when the trust-worthiness of the MC
        data in the helical phase itself is factored in.
        }
\label{fig:Hansmann_opar}
\end{center}
\end{figure}

\begin{figure}
\begin{center}
\epsfysize=7cm\epsfbox{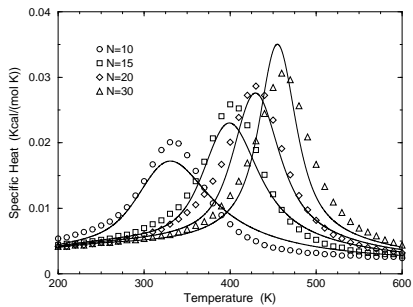}
\caption{
        Best fit to specific heat for model-water 2, obtained by
        simultaneous fitting 19 parameters to the cohort of MC data.
        A systematic fitting error can be seen, where the DCM (as
        presented  here) predicts too fast of an increase in the
        maximum peak as a function of chain length.
        }
\label{fig:Hansmann_c}
\end{center}
\end{figure}

The results of the simulated annealed best fits are given in Table
\ref{table:sip} for solvent independent DCM parameters, and Table
\ref{table:sdp} for solvent dependent DCM parameters.
Figures \ref{fig:Okamoto_opar} and \ref{fig:Okamoto_c} respectively show
the fit of helix content and specific heat for both gas phase and model-water
solvent 1. Figures \ref{fig:Hansmann_opar} and \ref{fig:Hansmann_c}
respectively show the fit of helix content and specific heat for all chain
lengths in model-water solvent 2. Good fits to helix content were achieved
for all 6 data sets, with the chain length of 30 in model-water solvent 2
showing greatest deviations in the helical phase. Likewise, the fits to
specific heat were in remarkably good quantitative agreement, considering
that the DCM parameters are taken as temperature independent over a 400 K
temperature range. Moreover, employing temperature dependent parameters
appears unnecessary for removing systemic error, because it can be
attributed to the over-simplified model of representing the peptide-solvent
hydrogen bonding as a {\em single} state. Overall, the minimalist network
rigidity model has successfully captured the essential physics that the MC
simulation does.

\section{Discussion\label{sec:discussion}}

The toy models in section \ref{sec:toy} and the helix-coil transition in section
\ref{sec:helix-coil} demonstrate how generic rigidity calculations are used
to construct a partition function at finite temperatures. Each framework in the
ensemble is weighted by a conformational degeneracy, $e^{\tau}$, that depends on
the type of constraints present and their specific placement relative to one
another. Effectively, the conformational degeneracy represents the free volume
available to a particular framework. It has long been recognized \cite{Grest}
that free volume plays an important role in both phase change and relaxation in
structural glasses. In the DCM, free volume is quantified by $\tau ({\cal F})$,
which depends on the strongest {\em independent} constraints that limit motion.
A direct connection between free volume and the degree of mechanical flexibility
is established through network rigidity --- an inherently long-range cooperative
interaction. Although the importance of rigidity in the conceptual understanding
of structural transitions is not new, the DCM allows the role of network rigidity
at finite temperatures to be calculated quantitatively.

\begin{table*}[b]
\begin{tabular}
[c]{c|c|c|c|c|c}\hline \hline
           & $U_0$             &       $V_c$      & $\delta_c$      & $V_c'$           & $\delta_c'$     \\ \hline
Gas        & -0.399            &     -0.321       &     3.603       &    -1.344        &   4.034         \\
           &  -0.67 $\pm$ 0.34 & -0.35 $\pm$ 0.08 & 3.58 $\pm$ 0.09 & -1.18 $\pm$ 0.22 & 3.62 $\pm$ 0.25 \\ \hline
Solvent 1  &   -1.154          &     -0.321       &     3.603       &    -3.095        &   3.523         \\
           & -1.40 $\pm$ 0.33  & -0.35 $\pm$ 0.08 & 3.58 $\pm$ 0.09 & -2.73 $\pm$ 0.18 & 3.55 $\pm$ 0.14 \\ \hline
Solvent 2  &   -0.399          &       -0.857     &     3.603       &    -3.095        &   3.523         \\
           &  -0.67 $\pm$ 0.34 & -0.87 $\pm$ 0.09 & 3.58 $\pm$ 0.09 & -2.73 $\pm$ 0.18 & 3.55 $\pm$ 0.14 \\ \hline \hline
\end{tabular}
\caption{List of 5 solvent dependent DCM parameters per solvent type. The
         same units and notation are used as in table \ref{table:sip}.
         \label{table:sdp}
         }
\end{table*}

In some respects the DCM is similar to a normal mode analysis in that
{\em entropies are additive over independent degrees of freedom}. If the 
system of interest can be well approximated as a network of coupled harmonic
oscillators, then the normal modes define an appropriate set of independent 
coordinates. However, normal mode analysis applied to the soft condensed phase
is subject to difficulties because of an-harmonic potentials \cite{Hayward}
that limit the range of validity over the assumed harmonic motions. In the
DCM, the ``strength'' of a constraint is inversely proportional to its free
volume quantified by a pure entropy. An extremely weak constraint having a
large free volume will pose no effective restrictions on conformational 
freedom. Although normal mode analysis is not intrinsically suited to deal
with bonds breaking and forming via thermal fluctuations, a self-consistent
phonon theory \cite{Cao} has been used to account for breaking and forming 
of hydrogen bonds in protein structure. Both the DCM and normal mode
analysis offer approximation schemes, but from opposite directions. For
example, soft an-harmonic (or flat) potentials are easier to deal with
in the DCM because they require less geometrical partitioning.

The DCM explicitly accounts for fluctuating topological constraints, allowing
a global picture to emerge in understanding structural self-organization. From
the three worked examples presented, it is seen that: 1) The effectiveness
of a constraint in changing the free energy of the system depends on
temperature and its location in relation to all other constraints. 2) Molecular
cooperativity derives from competition between frameworks having different
energetic and entropic contributions. More generally, a change in thermodynamic
conditions (temperature, pressure, pH, etc) can lead to a global re-arrangement
of optimally well placed constraints. 3) The most probable microstates will
often correspond to a characteristic pattern of constraints, manifesting itself
as structural self-organization. For example, in the helix-coil transition,
mechanical frameworks switch character as some constraint types tend to break
(alpha-helical torsion constraints and backbone hydrogen bond constraints)
while others tend to form (coil torsion constraints). This type of structural
self-organization has been produced in athermal network rigidity models
\cite{Thorpe1} applied to covalent glass networks, where redundant constraints
were suppressed to avoid strain energy. In other work to be published
elsewhere \cite{Jacobs_DCM2}, hydration effects are included in the DCM.
Structured water around a hydration site is considered to impose another type
of constraint on the peptide, where it is enthalpicly favorable and entropicly
unfavorable. Under certain thermodynamic conditions, cold denaturation occurs
as the character of constraint type and pattern change.

\subsection{The Helix-Coil Transition}

The helix-coil transition has been studied for nearly 50
years \cite{Schellman, Doig}.  For a simple statistical mechanical approach,
the Zimm and Bragg \cite{Zimm} (ZB) and Lifson and Roig \cite{Lifson}
(LR) models are commonly used. The ZB and LR models share two types
of parameters --- referred to as nucleation and propagation parameters.
Only two and three dimensional transfer matrices are required for the ZB
and LR models respectively \cite{Poland}. Without a doubt, the application
of the LR model to explain experimental data has been very fruitful over
the years. The question then arises, why use the more complicated DCM
when the traditional LR model will do?

The DCM clearly makes a distinction between a cooperative process
governing a structural transition to that of a non-cooperative process
that happens to have a sharp transition. A true signature for the degree
of cooperativity is in how the transition temperature depends on chain
length. The MC simulation data from Y. Peng {\em et. al.} \cite{Hansmann}
shows a large degree of cooperativity, as the transition temperature
dramatically increases by 130 K when increasing chain length from 10
to 30. The DCM is able to capture this degree of cooperativity without
requiring temperature or size dependent model parameters.

For comparison, the LR-model was also fitted to model-water solvent 2
MC data \cite{Hansmann}. LR relates the so called nucleation
parameter, $v$, and the propagation parameter, $w$, to partial
configurational integrals defined by coarse-grain sections of dihedral
angle space (helical or coil conformations) along the backbone. These
dimensionless parameters are expected to be functions of temperature,
where $-kT \ln v$ and $-kT \ln w$ represent microscopic component free
energies, and are treated phenomenologically \cite{u_parameter}. The
LR parameters can be written in a form similar to the DCM, where
$v = e^{2 \delta_v}$ and $w = e^{2 \delta_w} e^{- \beta V_w} $. Here
the parameters \{ $\delta_v$, $\delta_w$ and $V_w$ \} are taken as
temperature independent, and fitted to the 4 helix content curves. Note
that the $v$ parameter is assumed here to be temperature independent,
following common practice. Since the LR-model as commonly invoked does
not explicitly account for end effects, two additional parameters (not
model parameters) are required to account for helix content baselines.

\begin{figure}
\begin{center}
\epsfysize=7cm\epsfbox{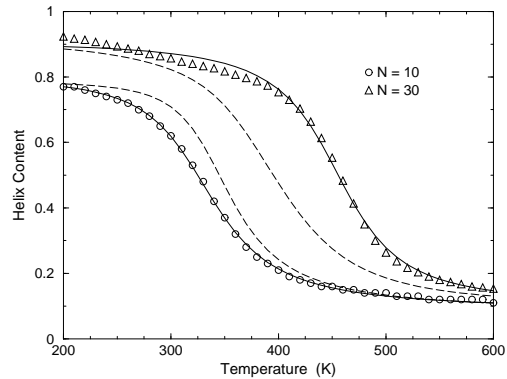}
\caption{
        The solid lines through the MC data for model-water
        solvent 2 show the best 5 parameter fit for each size chain
        separately using the standard LR-model. The dashed line on
        the (left, right) corresponds to the LR prediction of
        helix content for a chain of length (30, 10) using the
        best fit parameters obtained from chain length (10, 30).
        }
\label{fig:Lifson_fit}
\end{center}
\end{figure}

Helix content for chain lengths 10, 15, 20, and 30 were {\em individually}
fitted with the LR-model, each with 5 fitting parameters, requiring a
total of 20 parameters. Figure \ref{fig:Lifson_fit} shows the
simulation data for chain length 10 and 30, as well as the best fit
for each size. In addition, the prediction for helix content for chain
length (30, 10) using the fitted parameters from chain length (10, 30)
respectively are shown. The LR-model in its three
parameter form does a very good job in fitting to each helix content
curve. However, as Fig. \ref{fig:Lifson_fit} clearly shows, the
fitted parameters obtained for one size, cannot be used to predict
helix content of a different size. {\em The LR parameters are inherently
non-transferable because they depend on the size of the system.}
Although the sharpness of the helix content curve is accounted for
in the so called nucleation parameter, the mechanism creating the
cooperativity is completely missed in this simplest three-parameter form.
To be fair, a simultaneous fit to all four helix content curves was
attempted using 12 parameters (4 model parameters and 8 baseline
parameters). The extra LR-model parameter was introduced by letting
$v = e^{2 \delta_v} e^{- \beta V_v} $. Not surprising, no good
simultaneous fit solutions was possible.

Bierzynski and Pawlowski (BP) \cite{Bierzynski2} show that the nucleation
parameter is required to be a function of chain length due to the long range
character of helix formation. It seems unsporting to us to predict a helix
with parameters that vary with chain length. Furthermore, BP demonstrate
that a common implementation of the LR model predicts thermodynamic state
functions that are erroneously
path dependent: giving slightly different results depending on which end of
the peptide the computation begins at, and wrong predictions when prenucleated
peptides are considered. Fundamentally, the so called nucleation parameter is
ill-defined for use in calculating a partition function \cite{Bierzynski2},
and its wide spread use has created misconceptions \cite{nucleation}. The
DCM avoids these issues. The DCM has long range character through network
rigidity, thus recourse to length dependent parameters is unnecessary.

The DCM is actually very similar to the LR model. Both models are based
on parameters that can be derived from local microscopic free energies.
The difference is that the DCM attempts to include non-local cooperative
interactions explicitly by using generic rigidity calculations to account
for the non-additivity of entropy. Yet it is possible to construct a DCM
where there is very little entropic competition between constraint types,
such as given in column (a) in table \ref{table:dim}. In this case,
the DCM for a helix-coil transition is {\em identical} to the
{\em general form} of the LR model. It is worth noting that the
two commonly used LR parameters \cite{u_parameter}, ($v,w$), are only
a subset of sixteen parameters that must be defined for each possible
type of propagation (i.e. $aac$ $\rightarrow$ $aca$, and 15 more). Lifson
and Roig simplified the model considerably to solve it analytically.
Unfortunately, the advantages of simplifying the mathematical form of
the model has lead to non-transferability of parameters that have
created many inconsistencies in the literature \cite{Bierzynski1}. With
modern computers it is no longer necessary to invoke the two parameter
form of the LR model. The disadvantage of retaining the two parameter form
is that the parameters become strongly dependent functions of temperature
\cite{Okamoto,Hansmann,Scheraga} and chain length \cite{Okamoto,
Hansmann,Bierzynski2}.

\subsection{Solvent Effects on the Helix-Coil Transition}

\begin{figure}
\begin{center}
\epsfysize=7cm\epsfbox{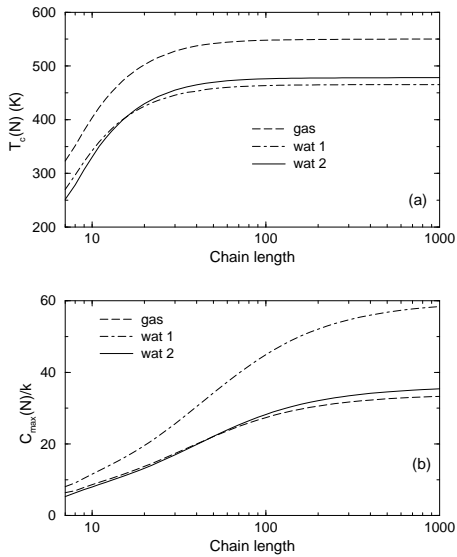}
\caption{The (a) transition temperature and (b) maximum value of the specific
         heat as a function of chain length for
         gas phase and model-water solvents 1 and 2. The parameters used in
         generating these curves are given in Tables \ref{table:sip} and
         \ref{table:sdp}.
        }
\label{fig:Tc_lines}
\end{center}
\end{figure}

The DCM parameters naturally divide into two categories that are
expected to be either weakly or strongly dependent on solvent conditions.
Moreover, the results obtained by fitting the DCM to MC simulation data
indicate the essential physics of the helix-coil transition for
poly-alanine is well described by the 10 solvent independent parameters
in table \ref{table:sip} and 5 solvent dependent parameters
given in table \ref{table:sdp}. For these DCM parameters
Fig. \ref{fig:Tc_lines} shows the
affect of solvent on the helix-coil transition. Comparing gas phase
and model-water solvents 1 and 2 with each other, we see that the
transition temperature and the sharpness of the transition can be
substantially modified. Not surprising, the gas phase transition
temperature is elevated with respect to model-water solvent, because
alternate hydrogen bonds from backbone to solvent cannot replace
intra hydrogen bonds as they break. The greater energy cost
to unravel the rigid helical structure requires a higher
transition temperature where gains in conformational entropy
can begin to compensate. It is also seen that the transition
temperature as a function of chain length for model-water solvents
1 and 2 are very similar, as one might expect if the differences
shown in Fig. \ref{fig:discrepancy} are viewed as systematic
uncertainties, rather than two different solvents.

\begin{figure}
\begin{center}
\epsfysize=7cm\epsfbox{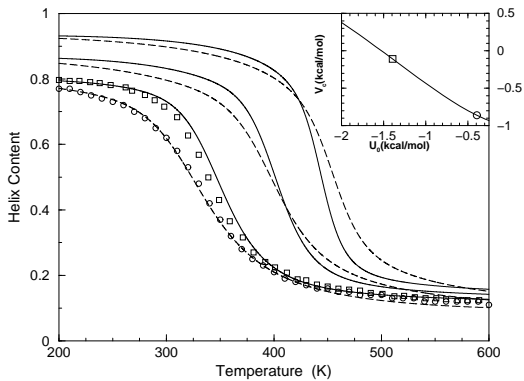}
\caption{
        Large graph: The dashed and solid curves show predictions for chain
        lengths 10, 15 and 30 that are obtained in a one-parameter best fit
        for $V_c$ when setting $U_o = -0.4$ and $U_o = -1.4$ respectively.
        The circles and squares show the results of MC simulation from
        Y. Peng {\em et. al.} \cite{Hansmann} and Okamoto \cite{Okamoto}
        respectively. Inset: The solid line shows the best fit value of
        $V_c$ along the ordinate as a function of $U_o$ along the abscissa.
        The (circle, square) indicates
        the $V_c$ and $U_o$ values used to generate the (dashed, solid)
        lines in the large graph. Due to the intrinsic uncertainty in the
        MC data, perhaps the best DCM parameter estimates are:
        $U_o=-0.900$ kcal/mol and $V_c=-0.485$ kcal/mol, which split
        the differences roughly in half.
        }
\label{fig:fuzz}
\end{center}
\end{figure}

The sharpness of the transition, as characterized by the maximum in
specific heat, is found to depend on the particular combination of
solvent dependent parameters. With respect to the gas phase, from
Fig. \ref{fig:Tc_lines} it is seen that the transition sharpens
considerably for model-water solvent 1, but remains virtually the
same for model-water solvent 2. These results correctly reproduce the
observations of the authors that generated the original MC simulation
data \cite{Okamoto, Hansmann}. Of course, model-water solvents 1 and 2
are actually the same, albeit systematic uncertainties shown in
Fig. \ref{fig:discrepancy}.  This 
uncertainty and the differences seen in Fig. \ref{fig:Tc_lines} are
the result of differences found in parameters ($U_o$ and $V_c$), listed 
in Table \ref{table:sdp}. Therefore, it is easy to interpolate
between the two different MC results within a 2-dimensional parameter
space. The interpolation was done by fitting only to model-water 
solvent 2 data. Letting $U_o$ range between $-1.4$ and $-0.4$ kcal/mol, 
a one parameter fit to obtain the optimal $V_c$ was performed, while 
holding $U_o$ and all other 17 parameters given in Tables 
\ref{table:sip} and \ref{table:sdp} fixed. It was found 
that the DCM model predictions smoothly change as a function of $U_o$.
In Fig. \ref{fig:fuzz}, the helix content is shown for model-water 
solvent, but now the uncertainties in the parameters $U_o$ and $V_c$ 
encompass both MC simulation results for the chain length of 10. 
Chain lengths of 10, 15 and 30 are shown in fig. \ref{fig:fuzz},
which gives some indication of the true uncertainties in helix
content for model-water solvent (using the ECEPP/2 force field).

In the DCM presented here, solvent effects on the helix-coil
transition were described well using just five parameters. A better
description is possible by including more states representing the
peptide to solvent interactions. In other work \cite{Jacobs_DCM2}
hydration constraints are included, for example. Furthermore,
inverted transitions from coil to helix as temperature increases
from low to high can be described.

\subsection{Molecular Cooperativity}

Admittedly, the DCM requires more effort than the LR model to describe the
helix-coil transition. The benefit of this additional labor is that the
final parameterization for understanding the nature of competing microscopic
interactions becomes considerably less complicated in the end.
In particular, the DCM offers the potential of having transferability
of parameters. Parameter transferability is intimately tied to the proper 
summation of component entropies, which is quantified in the DCM via the
{long-range underlying mechanical interaction between constraints}. From 
the fitted model parameters (given in Tables \ref{table:sip} and 
\ref{table:sdp}) it is seen from column (i) in table \ref{table:dim} that
a $244 \times 244$ transfer matrix was necessary to describe the MC
simulation results. The large size of the transfer matrix is an 
indication of a high degree of cooperativity among the hydrogen bonding
along the backbone.

In exchange for the non-transferable nucleation parameter to characterize 
the degree of cooperativity, it is characterized by a {\em rigidity
correlation length} in the DCM. The rigidity correlation length gives 
an indication of how far away from a point of interest that perturbations
in constraints will lead to little affect at the point of interest. It 
can be roughly estimated at the helix-coil transition by locating the
crossover point where the shift in transition temperature becomes small 
as chain length increases. From Fig. \ref{fig:Tc_lines}, the rigidity 
correlation length is estimated to be $\approx 40$ amino acids for
both gas and model-water solvents, also corresponding to the inflection 
point on the curves for maximum specific heat. The correlation length
is quite long considering that in 1-dimension thermal fluctuations
severely reduce the effectiveness of the long-range nature of network 
rigidity.

The primary motivation for introducing the DCM is to study flexibility and
stability in proteins \cite{Dang}. The concept of a rigidity correlation 
length applies to any topology of constraints, such as found in globular
proteins. The DCM can be used to directly study the affect of hydrogen bonds
on protein stability, which has been difficult to ascertain experimentally
and theoretically. Not only does the answer depend on the specific 
thermodynamic conditions, but also on the particular hydrogen bond in 
question. Stability questions are particularly difficult to answer when
there is a high degree of cooperativity in a molecular system. Proteins
are particularly interesting, where it has been suggested that the folding
pathway is encoded in the hydrogen bond network \cite{Rader,Hespenheide}.
In addition, mechanical stability probed by single-molecule force spectroscopy
appears to depend on the kinetic stability of the hydrogen bond network
\cite{Carrion} --- also a cooperative process that can be addressed 
within a DCM. More generally, the DCM describes protein folding as
a manifestation of a structural self-organization caused by the 
topological optimization of constraint placement. Indeed, all model
calculations presented here suggest that the most probable frameworks
correspond to well defined structural units (such as secondary 
structure, protein domains, etc) that change character under different 
thermodynamic conditions.

\section{Conclusion\label{sec:conclusion}}

The DCM generalizes the $T=0$ generic rigidity calculation to finite
temperatures by quantifying constraints with energetic and entropic 
characteristics. The effectiveness of a constraint strongly depends 
on its type and where it is placed in the network in relation to all 
other constraints. Generic rigidity is then used as an underlying long-range
mechanical interaction between constraints, providing the mechanism for 
the non-additive property of component entropies. The DCM accounts for 
fluctuating topological patterns of constraint placements. From a 
computational point of view, the network rigidity calculations are easy 
to implement by invoking fast graph-algorithms that are available in 
two-dimensions \cite{Jacobs0,Jacobs1} for general networks, and in 
three dimensions \cite{Jacobs3} for bond-bending networks.

In this paper, a DCM applied to the helix-coil transition was considered
in detail and compared to the Lifson Roig model. Thermodynamic state
functions are calculated exactly, without recourse in using a
nucleation parameter. The helix-coil transition in peptides
is special only in that it can be exactly solved as a 1-dimensional system
using a transfer matrix method. Our use of the DCM has been to coarse grain
into the smallest number of states necessary to describe the physics at hand.
For example, alpha helix and coil backbone states are used in modeling the
helix-coil transition. In this work, 12 different thermodynamic response 
functions where described well by the DCM using 20 parameters that are
independent of temperature and chain length. The entropic parameters 
indicate that the degree of cooperativity extends over approximately 
40 amino acids. 

As a practical application, the DCM may be able to predict helix formation 
in proteins with parameters derived from helix-coil transition studies.
The DCM is readily scalable to include more types of interactions,
where far more backbone states could have been introduced such as
3-10 helix, beta sheet, beta turn, hydrated or not hydrated, buried or 
surface exposed. If the DCM parameters are found to be transferable (as
we expect) flexibility and stability studies on proteins will be far more
feasible, because the DCM gets more physics out with fewer parameters.
The DCM has the potential to gain a better understanding of these issues
from a mechanical point of view. More generally, the DCM gives a description
of a coarse graining procedure to describe physical systems. Its 
applicability goes beyond biopolymers, offering a new paradigm not 
previously available.

\begin{acknowledgments}
The authors are grateful for financial support from California
State University, Northridge, the Research Corporation (CC5141)
and to the NIH (GM48680-0952). We thank Dennis Livesay for
many useful discussions. We also thank Prof. Hansmann 
sharing their raw MC simulation data with us.
\end{acknowledgments}



\begin{thebibliography}{99}
\bibitem{footnote1} The entropy loss from a redundant constraint is expected to be
                    a small fraction of what would occur if the constraint was
                    independent. Accounting for this second order effect is
                    possible at the expense of losing simplicity.
\bibitem{footnote2} The DCM explicitly neglects relaxation within constraints. 
                    Relaxation in structure can be described by incorporating
                    more geometrical bins at the expense of losing simplicity.
\bibitem{prop_test} The propagation rule was applied from left to right and right 
                    to left on 200,000 chains with length of 100,000 as well as
                    millions of chains of lengths ranging from 10 to 1000 all 
                    having random conformations. These results were compared
                    to exact calculation using the pebble game per framework. 
                    Agreement between all three results was always obtained.
\bibitem{bound_s}   Technically adding the $s$-conformations, rank $r_s$ 
		    and the special matrices \{$S$, $R$, $Q$\} required expanding
		    the dimension of the direct product space. The $Dim(T)$ referes to 
		    the subspace that excludes the $s$-state.
\bibitem{best_fit} The objective of finding an optimal set of parameters with available 
                   data in the literature is out of scope of this work. The subject of 
                   obtaining an accurate DCM parameterization is deferred to future work.
\bibitem{u_parameter} A third parameter called $u$ in the LR-model is arbitrarily
                      set to unity, which is analogous to the arbitrary choices
                      made in fixing some DCM parameters.
\end{thebibliography}
\end{document}